\documentclass{iopart}

\usepackage{iopams}
\usepackage{graphicx}
\usepackage{xcolor}
\usepackage[square,sort&compress,numbers]{natbib}
\usepackage[colorlinks,linkcolor=blue!80!black,urlcolor=purple!50!black,citecolor=green!50!black]{hyperref}

\usepackage{mathrsfs}

\newcommand{\be}{\begin{equation}}
\newcommand{\ee}{\end{equation}}
\newcommand{\bea}{\begin{eqnarray}}
\newcommand{\eea}{\end{eqnarray}}

\newcommand{\eqref}[1]{\mbox{Eq.~(\ref{#1})}}

\makeatletter

\newcommand{\Rmnum}[1]{\expandafter\@slowromancap\romannumeral #1@}
\makeatother

\begin{document}
	
\title{Condensed Matter Physics in Time Crystals}

\author{Lingzhen Guo$^{1}$ and Pengfei Liang$^{2,3}$}
\address{$^1$Max Planck Institute for the Science of Light (MPL), Staudtstrasse 2, 91058 Erlangen, Germany}
\address{$^2$Beijing Computational Science Research Center, 100193 Beijing, China }
\address{$^3$Abdus Salam ICTP, Strada Costiera 11, I-34151 Trieste, Italy }
\ead{lingzhen.guo@mpl.mpg.de}

\begin{abstract}
Time crystals are physical systems whose time translation symmetry is spontaneously broken. Although the spontaneous breaking of continuous time-translation symmetry in static systems is proved impossible for the equilibrium state, the discrete time-translation symmetry in periodically driven (Floquet) systems is allowed to be spontaneously broken, resulting in the so-called {\it Floquet} or {\it discrete time crystals}. While most works so far searching for time crystals focus on the symmetry breaking process and the possible stabilising mechanisms, the many-body physics from the interplay of symmetry-broken states, which we call the {\it condensed matter physics in time crystals}, is not fully explored yet. This review aims to summarise the very preliminary results in this new research field with an analogous structure of condensed matter theory in solids. The whole theory is built on a hidden symmetry in time crystals, i.e., the {\it phase space lattice symmetry}, which allows us to develop the band theory, topology and strongly correlated models in phase space lattice. In the end, we outline the possible topics and directions for the future research.   
\end{abstract}

\maketitle

\tableofcontents
\title[Condensed Matter Physics in Time Crystals]
\maketitle

\section{Brief introduction to time crystals}

\subsection{Wilczek's time crystal }
The idea of time crystals was proposed by Frank Wilczek in 2012 \cite{wilczeck2012prl-1,wilczeck2012prl-2}. He raised the question that whether {\it continuous time-translation symmetry (CTTS)} might be spontaneously broken in a {\it closed time-independent} quantum mechanical system, namely, some physical observable of the quantum-mechanical ground state exhibits persistent periodic oscillation. Wilczek also noticed that the Heisenberg equation of motion for an operator, i.e., 
\bea
\langle \Psi_E | \dot{\mathcal{O}}|\Psi_E \rangle=i\langle \Psi_E | [H,\mathcal{O}]|\Psi_E \rangle=0
\eea
for any eigenstate $\Psi_E$ of $H$, seems to make the possibility of quantum time crystals implausible. Nevertheless, he proposed a model consisting of particles with contact attractive interaction on a ring threaded by an Aharonov-Bohm (AB) flux, and found that this model can have the ``ground state'' of  a {\it rotating soliton}, which satisfies the definition of time crystals. 

However, the spontaneous breaking of CTTS is against the energy conservation law. The rotating soliton generates a time-dependent force field on the ring. Imagine that we put a particle at some point on the ring when the force exerted by the soliton is weak. As the soliton is approaching the point, the particle will be accelerated and absorb some energy from the rotating soliton.  In this way, some positive net energy is produced without lowering the energy of the rotating soliton as the ground state has the lowest energy by definition. Wilczek's ring is actually a new version of perpetual motion machine, which implies there must be something wrong in his analysis. 
In the history of science, although all the proposals of perpetual motion machines failed, several famous ones, e.g.,  Maxwell's demon \cite{leff2002book} and Brownian ratchet \cite{feynman1963book}, played significant roles in advancing science and still inspire new work today. Disproving Wilczek's idea needs serious and rigorous mathematical proof, which may deepen our understanding of physics laws.

\subsection{No-go theorem }
Bruno first pointed out Wilczek's rotating solution is not the true ground state for the model he proposed \cite{bruno2013prlcomment}. The correct ground state is actually a stationary state without breaking CTTS \cite{kanamoto2003pra}.
However, it is too early to disprove Wilczek's general idea of time crystals by precluding his soliton model. In the original work \cite{wilczeck2012prl-1}, Wilczek started his analysis from considering a superconducting ring threaded by a magnetic flux that is a fraction of the flux quantum. If Cooper pairs are free of interaction, the ground state is indeed a constant uniform superconducting current along the ring. But if the Cooper pairs have some kind of interaction (not only restricted to the contact form), one can ask whether a charge density wave (or Wigner crystal) can appear and rotate as the ground state. Actually, this rotating change density wave was soon proposed with trapped ions after Wilczek's original work \cite{li2012prl}. However, a detailed analysis showed that the rotating charge density wave cannot be the ground state \cite{Nozieres2013epl}. A ``no-go theorem'' was proved later by Bruno \cite{bruno2013prl}, which rigorously rules out the possibility of spontaneous ground-state (or thermal-equilibrium) rotation for a broad class of systems.

Spontaneous symmetry breaking occurs in the thermodynamic limit, where the system size $N$ (e.g., the number of particles in Wilczek's model or the number of spins in Ising or Heisenberg models) goes to infinity. A general method to treat the thermodynamic limit correctly was given by Bogoliubov \cite{bogoliubov1970book}: first calculating physical quantities for the system with finite size $N$ and perturbative symmetry breaking potential $v$, then taking the limit $N \rightarrow \infty$ before taking the limit $v\rightarrow 0$. Following Bogoliubov's prescription, Bruno found that, in the presence of an external symmetry-breaking potential rotating at angular velocity $\omega$,  the $n$th-level energy of the system in the static frame is 
\bea
E^{(\omega)}_{\phi,n}=E^{(0)}_{\phi,n}+\frac{1}{2}I_{\phi,n}\omega^2+O(\omega^3),
\eea 
where $\phi$ is the flux threaded through Wilczek's ring and $I_{\phi,n}$ is the moment of inertia for the $n$th energy level of the system. Bruno's crucial finding is that $I_{\phi,0} >0$ for the ground state breaking the rotational symmetry and $I_{\phi,0}=0$ otherwise. Furthermore, he proved the inequality $E^{(\omega)}_{\phi,n}>E^{(0)}_{\phi,n}$ for any finite value of $\omega \neq 0$. For the system in the thermal equilibrium at inverse temperature $\beta$, the same conclusion holds for the free energy, i.e., $\mathcal{F}^{(\omega)}_{\phi,\beta}>\mathcal{F}^{(0)}_{\phi,\beta}$ for $\omega \neq 0$. These two inequalities consist of the ``no-go theorem", which strictly prohibits the existence of spontaneously rotating time crystals for the ground state and thermal equilibrium state. Bruno also pointed out that
 the moment of inertia for an excited state $I_{\phi,n}(n>0)$ may be negative, which allows for  the existence of time crystal at some excited eigenstate \cite{sacha2017prl,autti2018prl} or time-crystal Sisyphus dynamics near the energy minimum \cite{shapere2019pnas}.
 
However, as Bruno's proof is restricted to a special model Hamiltonian, Wilczek and others later proposed new types of time crystals to avoid Bruno's no-go theorem \cite{wilczek2013prl,yoshii2015prb}. A more general no-go theorem was given by Watanabe and Oshikawa (WO) \cite{WO2015prl}. Using the local order parameter $\hat{\phi}(\vec{x},t)$, WO presented a precise definition for the time crystals, that is, the system is a time crystal if the correlation function
\bea
\lim_{V\rightarrow\infty} \langle e^{i\hat{H}t}\hat{\phi}(\vec{x},0)e^{-i\hat{H}t}\hat{\phi}(0,0)\rangle\rightarrow f(t)
\eea
for large $|\vec{x}|$ (much greater than any microscopic scales) exhibits a nontrivial periodic oscillation in time $t$. In terms of the integrated order parameter $\hat{\Phi}\equiv\int_Vd^dx\hat{\phi}(\vec{x},0)$, the condition reads equivalently $\lim_{V\rightarrow\infty} \langle e^{i\hat{H}t}\hat{\Phi}e^{-i\hat{H}t}\hat{\Phi}\rangle/V^2\rightarrow f(t)$. At zero temperature, WO was able to prove the following equality for the ground state 
\bea
 \frac{1}{V^2}\Big|\langle 0| \hat{\Phi}e^{-i(\hat{H}-E_0)t}\hat{\Phi}|0\rangle-\langle 0|\hat{\Phi}^2|0\rangle \Big|\leq C\frac{t}{V},
 \eea
where $E_0$ is the ground state energy and $C$ is a constant dependent on $\hat{\Phi}$ and $H$ but not on $t$ and $V$. The above inequality holds for the interaction which decays exponentially as a function of the distance or decays as power-law $r^{-\alpha}(\alpha>0)$. Therefore, for any fixed time $t$, the correlation function $f(t)$ remains constant in the limit $V\rightarrow \infty$. The conclusion can be extended to the case of canonical ensemble at finite temperature by the help of Lieb-Robinson bound \cite{lieb1972cmp}.
%
%
A related definition of time translation symmetry breaking (TTSB) was discussed in the language of $C^*$ algebras \cite{khemani2017prb}, which recovers WO's result for the absence of TTSB in equilibrium systems. It is still an open question that if one can bypass the limitations imposed by WO's no-go theorem and reach the desirable phase of a quantum time crystal in the ground state of an isolated quantum system without a driver \cite{kozin2019prl}.

\subsection{Discrete time-translation symmetry breaking}\label{sec-ftc}
The no-go theorem valid for equilibrium state does not exclude the possibility of spontaneous TTSB in nonequilibrium states, e.g., the intrinsically nonequilibrium setting of a periodically driven (Floquet) system, whose Hamiltonian has the {\it discrete time-translation symmetry (DTTS)} such that $H(t)=H(t+T_d)$ with $T_d=2\pi/\omega_d$ the period of driving field. It was found theoretically that the DTTS can be spontaneously broken in periodically driven atomic systems \cite{sacha2015pra} and spin chains \cite{khemani2016prl,else2016prl,von2016prb,yao2017prl}. 

The DTTS is said to be broken if for every ``short-range correlated" state $|\psi\rangle$, there exists an operator $\hat{\Phi}$ and integer $n>1$ such that 
\bea\label{psinT}
\langle\psi(nT_d)|\hat{\Phi}|\psi(nT_d)\rangle=\langle\psi(0)|\hat{\Phi}|\psi(0)\rangle,
\eea
namely, the system responds at a fraction $\omega_d/n$ of the original driving frequency \cite{else2016prl}. Here, a state $|\psi\rangle$ is said to be ``short-range correlated" if, for any local operator $\hat{\phi}(\vec{x})$, 
\bea
\langle \psi | \hat{\phi}(\vec{x})\hat{\phi}(\vec{x}')|\psi \rangle \rightarrow \langle \psi | \hat{\phi}(\vec{x})|\psi \rangle\langle \psi |\hat{\phi}(\vec{x}')|\psi \rangle 
\eea
as $|\vec{x}-\vec{x}'|\rightarrow \infty$ \cite{else2016prl}.
%
%
We call the robust short-ranged correlated state with broken DTTS the {\it Floquet time crystal} (FTC) or {\it discrete time crystal} (DTC).

The Floquet time crystals have been realised in various experimental platforms, e.g., trapped ions \cite{zhang2017nature}, diamond nitrogen vacancy centers \cite{choi2017nature}, nuclear spins \cite{rovny2018prl,rovny2018prb,pal2018prl}, superfluid $^3$He-B (magnon BEC) \cite{autti2018prl} and ultracold atoms \cite{smits2018prl}. 
For more recent progress on the study of time crystals, we refer the readers to other review papers \cite{sacha2017rpp,khemani2019brief,else2019discrete} and the references therein.

For the detailed description of Floquet time crystals, it needs some basics for the periodically driven systems, i.e., the Floquet theory. In the next section, we will first introduce the Floquet theory and Magnus expansion, then describe in detail how to realise the Floquet time crystals by breaking DTTS.





\section{Floquet time crystals}

\subsection{Floquet theory}\label{sec-fbt}

The DTTS of Floquet systems with periodic Hamiltonian $H(t)=H(t+T)$ enables us to use the Floquet formalism \cite{Floquet1883,ince1956book,shirley1965pr,magnus1979book,grifoni1998pr}. We write the Schr\"odinger equation for the quantum system as $(\hbar=1)$
\bea\label{hxt}
\mathscr{H} (x,t)|\psi(x,t)\rangle=0 \ \ \ \ \mathrm{with}\ \ \ \ \mathscr{H} (x,t)\equiv H(x,t)-i\frac{\partial}{\partial t},
\eea 
where $x$ represents coordinates or spins. Similar to the Bloch theorem in solid state physics, the time periodicity of operator $\mathscr{H} (x,t)=\mathscr{H} (x,t+T)$ enforces the eigensolutions of Eq.~(\ref{hxt}) in the form of (so-called Floquet solution \cite{Floquet1883,ince1956book,shirley1965pr,magnus1979book,grifoni1998pr}) 
\bea\label{psialphat}
|\psi_\alpha(x,t)\rangle=e^{-i\epsilon_\alpha t}|\varphi_\alpha(x,t)\rangle \ \ \ \mathrm{with}\ \ \  |\varphi_\alpha(x,t)\rangle=|\varphi_\alpha(x,t+T)\rangle.
\eea
Here, $ |\varphi_\alpha(x,t)\rangle$ is called the {\it Floquet mode} and $\epsilon_\alpha$ is called the {\it quasienergy}, which is a real-valued variable modulo $\Omega=2\pi/T$. The terminology quasienergy is analogous to the quasimomentum characterizing the Bloch eigenstates in
a crystalline solid. By substituting Eq.~(\ref{psialphat}) into Eq.~(\ref{hxt}), we find
\bea\label{hxtvar}
\mathscr{H} (x,t)|\varphi_\alpha(x,t)\rangle=\epsilon_\alpha|\varphi_\alpha(x,t)\rangle.
\eea 
According to Eq.~(\ref{psialphat}), the new state $|\varphi_{\alpha,n}(x,t)\rangle\equiv e^{in\Omega t}|\varphi_{\alpha}(x,t)\rangle$ ($n=0,\pm 1, \pm 2,\cdots$) is also the Floquet mode but with the shifted quasienergy $\epsilon_{\alpha, n}\equiv \epsilon_\alpha + n\Omega$. Therefore, we can map all the quasienergies into a first Brillouin zone, e.g., $0\leq\epsilon_\alpha < \Omega$.

In order to calculate the quasienergy
spectrum and the Floquet eigenstates
of $\mathscr{H}(x,t)$, it is convenient to introduce a
composite Hilbert space $\mathscr{R}\otimes\mathscr{T}$, where
$\mathscr{R}$ is the spatial (or spin) space and $\mathscr{T}$ is the space of functions with time periodicity
$T$.  We can choose the time-{\it independent} states
$|\phi_m(x)\rangle$ satisfying $\langle \phi_n(x)|\phi_m(x)\rangle=\delta_{nm}$ as the orthonormal basis of space
$\mathscr{R}$ and the time-{\it dependent} Fourier vectors $|e^{iq\Omega t}\rangle$ with
$q=0,\pm1,\pm2,\ldots,$ as the orthonormal basis of space
$\mathscr{T}$. Then, we can obtain the matrix of  $\mathscr{H}(x,t)$ in the complete set of $\{|\phi_m(x)\rangle\otimes |e^{iq\Omega t}\rangle\}$ and calculate its quasienergy
spectrum and the Floquet eigenstates.

On the one side, according to Eq.~(\ref{psialphat}), the dynamics of the Floquet eigenstate $|\psi_\alpha(x,t)\rangle$ at stroboscopic time moment $t=nT$ ($n\in\mathbb{Z}$) is given by
\bea\label{psint1}
|\psi_\alpha(x,nT)\rangle=\big(e^{-i\epsilon_\alpha T}\big)^n|\varphi_\alpha(x,0)\rangle.
\eea
On the other side, from the Schr\"odinger equation, the time evolution of the Floquet eigenstate $|\psi_\alpha(x,t)\rangle$ can be directly written in the Dyson form
\bea\label{psint2}
|\psi_\alpha(x,nT)\rangle=\Big(\mathcal{T}e^{-i\int_{0}^{T}H(t)dt}\Big)^n|\psi_\alpha(x,0)\rangle,
\eea
where $\mathcal{T}$  is the time-ordering operator. Comparing Eq.~(\ref{psint1}) to Eq.~(\ref{psint2}), the stroboscopic dynamics of Floquet system can be described by an effective time-independent Hamiltonian $H_F$ determined from 
\bea\label{uf}
e^{-iH_FT}=\mathcal{T}e^{-i\int_{0}^{T}H(t)dt}\equiv U_f
\eea
or written formally as
\bea\label{HF}
H_F=i\frac{1}{T}\ln\Big(\mathcal{T}\exp\big[-i\int_{0}^{T}H(t)dt\big]\Big)
\eea
which has the eigenvalues $\epsilon_\alpha$ and eigenstates $|\varphi_\alpha(x,0)\rangle$. The time-independent Hamiltonian $H_F$ is called {\it Floquet Hamiltonian}.
%
%

\subsection{Magnus expansion}
It is tempting to calculate the explicit form of the Floquet Hamiltonian (\ref{HF}). However,  except very few examples like a driven harmonic oscillator, it is impossible to obtain a closed form of $H_F$ in general. In the regime where the driving frequency is much larger than the characteristic frequency of the system, the Floquet Hamiltonian can be given by the so-called {\it Floquet-Magnus expansion}. 

Given the time differential equation of motion for the time evolution operator $i\partial_t U(t,t_0)=H(t)U(t,t_0)$, the Magnus theorem \cite{Magnus954zz,blanes2009pr} claims that the time evolution operator can be written in the form of $U(t,t_0)= \exp [-i\Omega (t,t_0)]$ with the operator $\Omega(t,t_0)$ determined by the following differential equation
\bea\label{ddtomega}
\frac{d}{dt}\Omega(t,t_0)=\sum_{n=0}^{\infty}\frac{B_n}{n!}[\Omega^{(n)},H],
\eea
where $[\Omega^{(n)},H]$ is the multiple commutator defined via the recursion relationship $[\Omega^{(n)},H]=[\Omega,[\Omega^{(n-1)},H]]$ with $[\Omega^{(0)},H]=H$, and $B_n$ are Bernoulli numbers given by the identity $\frac{x}{e^x-1}=\sum_{n=0}^\infty\frac{B_n}{n!}x^n$. The first few nonzero Bernoulli numbers are $B_0=1,\ B_1=-1/2,\ B_2=1/6,\ B_4=-1/30$ and $B_{2m+1}=0$ for $m\geq 1$. 

We now replace $H$ by $\varepsilon H$ in Eq.~(\ref{ddtomega}) and try a solution in the form of the Magnus series
\bea\label{omegasum}
\Omega(t,t_0)=\sum_{n=1}^{\infty}\varepsilon^n\Omega_n(t,t_0).
\eea
By substituting the above series (\ref{omegasum}) into Eq.~(\ref{ddtomega}) and equating powers of $\varepsilon$, we can obtain in principle all the expansion orders $\Omega_n(t)$. For example, the first three orders are given by \cite{blanes2009pr} 
\bea
\Omega_1(t,t_0)&=&\int_{t_0}^tdt_1H(t_1), \nonumber\\
 \Omega_2(t,t_0)&=&\frac{1}{2}\int_{t_0}^tdt_1\int_{t_0}^{t_1}dt_2[H(t_1),H(t_2)], \nonumber\\
 \Omega_3(t,t_0)&=&\frac{1}{6}\int_{t_0}^tdt_1\int_{t_0}^{t_1}dt_2\int_{t_0}^{t_2}dt_3\big([H(t_1),[H(t_2),H(t_3)]]\nonumber\\
 &&+[[H(t_1),H(t_2)],H(t_3)]\big).
\eea
For the explicit expressions of higher orders, one can find in Refs.~\cite{BialynickiBirula1969wn,annales1970,SAENZ2002357,STRICHARTZ1987320,suarez2001jomp}.

We apply the above Magnus theorem to Floquet system with periodic Hamiltonian $H(t)$ in the form of Fourier harmonics
\bea
H(t)=\sum_{l\in \mathbb{Z}}H_le^{il\Omega t}\ \ \ \mathrm{with} \ \ \ \Omega=\frac{2\pi}{T}.
\eea
By setting the time evolution interval $t\in [0,T]$, we have the Floquet Hamiltonian $H_F=T^{-1}\Omega (T,0)$ from Eq.~(\ref{HF}) and the Magnus theorem. Letting the perturbative parameter $\varepsilon=1$ in Eq.~(\ref{omegasum}), we decompose the Floquet Hamiltonian in the form of Floquet-Magnus expansion $H_F=\sum_{n=0}^\infty H^{(n)}_F$ with the first three orders given by \cite{bukov2015aip}
\bea\label{h0h1h3}
H^{(0)}_F&=&H_0=\frac{1}{T}\int_{0}^TdtH(t), \nonumber\\
 H^{(1)}_F&=&\frac{1}{\Omega}\sum_{l=1}^\infty\frac{1}{l}[H_l,H_{-l}], \nonumber\\
 H^{(2)}_F&=&\frac{1}{\Omega^2}\sum_{l\neq0}\Bigg(\frac{[H_{-l},[H_0,H_{l}]]}{2l^2}+\sum_{l'\neq 0,l}\frac{[H_{-l'},[H_{l'-l},H_{l}]]}{3ll'}\Bigg).
\eea
The leading term $H^{(0)}_F=H_0$ is nothing but the time average of the original Hamiltonian over one time period $T$.

We should emphasise that, although the Magnus expansion method has been used for a long history in physics, the precise radius of convergence of Magnus series is still an open mathematical problem. A sufficient condition for the convergence of Magnus series (\ref{omegasum}) in the time interval $t\in[0,t_c]$ is given by \cite{blanes2009pr}
$
\int_0^{t_c}\|H(t)\|dt <\pi,
$
where the norm $\|A\|$ is defined as the square root of the largest eigenvalue of $A^\dagger A$, and the operator $H(t)$ needs to be a bounded operator in Hilbert space. More precise criterion of convergence can be found in Ref.~\cite{Casas2007jpa}. For the Flouqet-Magnus expansion, the convergence condition becomes $\int_0^{T}\|H(t)\|dt <\pi$ with $T=2\pi/\Omega$ the time period. However, this convergence condition is not guaranteed in physical systems. For instance, since the generic nonintegrable infinite-size systems eventually heat up to infinite temperature  for an arbitrary
driving frequency, it was concluded that the radius of convergence for the Floquet-Magnus expansion  vanishes in the thermodynamic limit and the quantum ergodicity results in the divergence of the
expansion \cite{dalessio2013aop,lazarides2014pre,dalessio2014prx,ponte2015aop,Moessner2017np,bukov2016prb}. It is still unclear whether the divergence of the Floquet-Magnus expansion necessarily results in the heat up to infinite temperature due to the difficulty to estimate the radius of convergence for nonintegrable many-body systems. It was recently conjectured that the divergence of the Floquet-Magnus expansion is a universal feature of periodically driven nonlinear systems with energy localization \cite{haga2019pre}. Nevertheless, since the system costs exponentially long time to reach the infinite-temperature featureless state if the driving frequency is much larger than the local energy scale, the prethermal dynamics can be well described by a truncation of the Floquet-Magnus expansion \cite{goldman2014prx,mori2016prl,kuwahara2016aop,bukov2016prb,abanin2017cmp,abanin2017prb,ho2018prl,machado2019prr}.

\subsection{Subharmonic modes}

\begin{figure}
  \includegraphics[width=\linewidth]{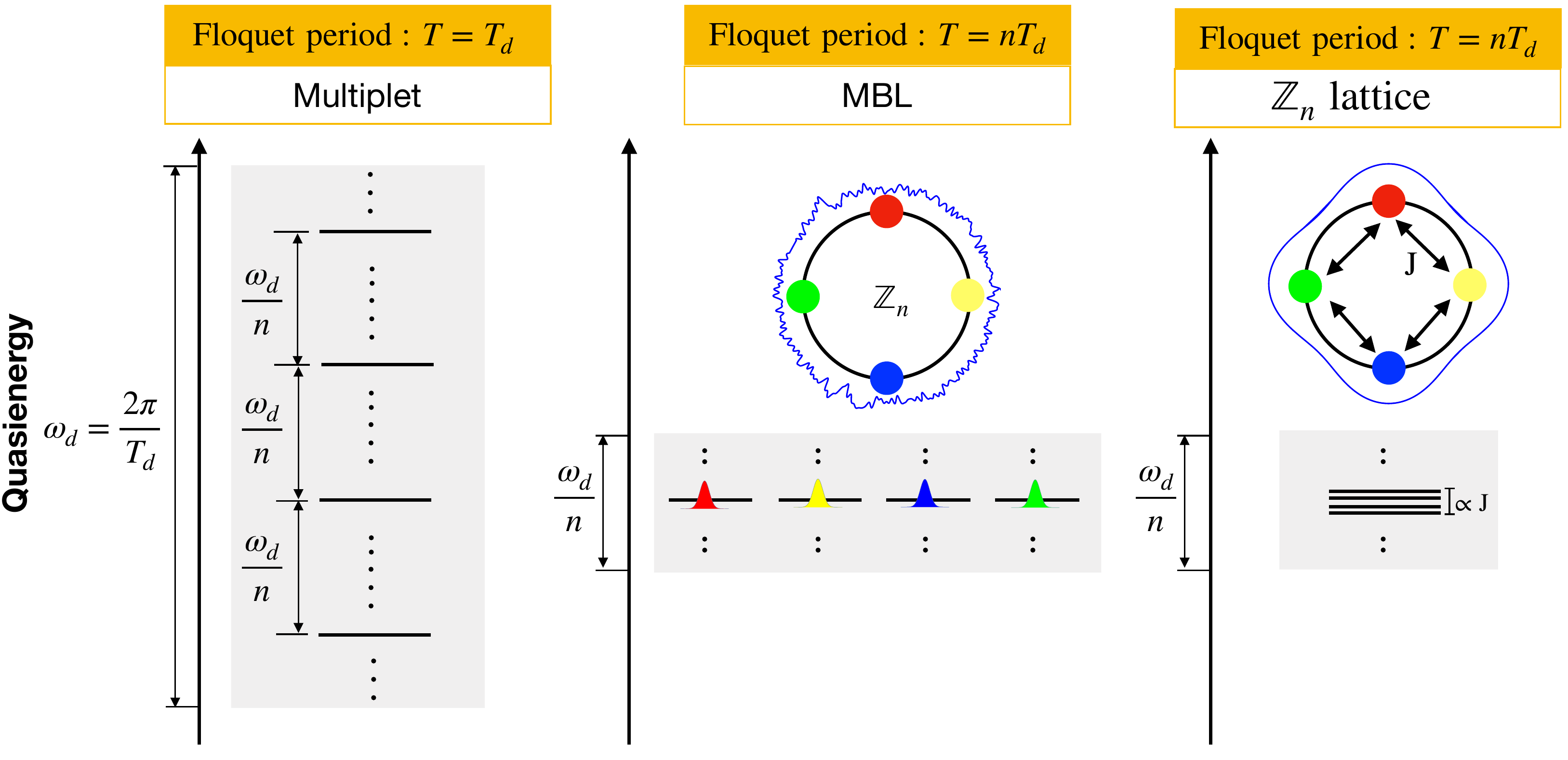}
  \caption{{\bf Floquet time crystals.} (Left) A $n$-component multiplet of Floquet eigenstates emerges in the quasienergy spectrum with the choice of Floquet period equal to the driving period $T=T_d$.
  (Middle) The $n$ Floquet multiplets become degenerate (i.e., $\mathbb{Z}_n$ symmetry) with the choice of Floquet period $T=nT_d$. In the thermodynamic limit, the $\mathbb{Z}_n$ symmetry of Floquet eigenstates are spontaneously broken into subharmonic modes (coloured wave packets), which can be stabilised by many body localisation (MBL) with disorders (blue noisy curve). (Right) If there is an additional $\mathbb{Z}_n$-lattice potential (blue curve), the subharmonic modes tunnel to its nearest neighbour with rate $J$ resulting a Floquet band with width $\propto J$. In the limit of $J\rightarrow 0$, the Floquet eigenstates are broken into robust subharmonic modes localised on the $\mathbb{Z}_n$-lattice sites.
  }
  \label{Fig-CMPinFTC}
\end{figure}

To understand the formation of Floquet time crystals, let us consider a one dimensional chain of $N$ spin-half particles with a binary driving protocol
\bea\label{ht}
H(t)=\left\{\begin{array}{l}
H_1\equiv (1-\delta)B^x\sum_{i}\sigma^x_i, \ \ \ \ \ \ \ \ \ \ \ \ \ \ 0\leq t<\tau \hbar/B^x
 \\
H_2\equiv \sum_{i}J\vec{\sigma}_i\cdot\vec{\sigma}_{i+1}+B^z_i\sigma^z_i,\ \   \ \tau \hbar/B^x\leq t<T_d,
 \end{array}\right.
\eea 
where $\vec{\sigma}_i$ are Pauli operators, $B^z_i\in [0,W]$ is a random longitudinal field and $\delta$ is the imperfection of the uniform transversal field $B^x$. Here, we set the dimensionless time interval $\tau=\frac{\pi}{n}$ with $n\in \mathbb{N}$ and $T_d>\tau \hbar/B^x$ is the driving period. For the case of perfect driving ($\delta=0$) and clean system (disorder strength $W=0$), the Floquet unitary (\ref{uf}) has the form ($\hbar=1$) 
\bea
U_f
=e^{-i\big(T_d-\tau\big) \sum_{i}J\vec{\sigma}_i\cdot\vec{\sigma}_{i+1}}R_\tau \ \ \ \ \ \mathrm{with}\ \ \ \ \ R_\tau\equiv e^{-i\tau\sum_i\sigma^x_i},
\eea
where we have chosen the Floquet period $T=T_d$. 
Obviously, the short-range correlated state $|\phi_0\rangle\equiv|\uparrow\uparrow\uparrow\cdots\uparrow\rangle$ is NOT the eigenstate of $U_f$ since the operator $R_\tau$ rotates all the spins. 
The real Floquet eigenstates $|\psi_m\rangle$ ($m=0,1,\cdots,n-1$) can be constructed in the following form 
\bea\label{psim}
|\psi_m\rangle\propto\sum^{n-1}_{q=0}e^{i2mq\tau}R^q_\tau |\phi_0\rangle\equiv \sum^{n-1}_{q=0}e^{i2mq\tau}|\phi_q\rangle, 
\eea 
where $|\phi_q\rangle\equiv R^q_\tau |\uparrow\uparrow\uparrow\cdots\uparrow\rangle$ ($q=0,1,\cdots,n-1$). The Floquet eigenstates (\ref{psim}) have $\mathbb{Z}_n$ symmetry, i.e., $R^n_\tau |\psi_m\rangle=|\psi_m\rangle$ from $R_\tau |\psi_m\rangle=e^{-i2m\tau}|\psi_m\rangle$. Since the interaction part in $H_2$ is invariant by the unitary transformation of operator $R_\tau$, we have the quasienergy of Floquet eigenstate $|\psi_m\rangle$ from $U_f|\psi_m\rangle=e^{-i\epsilon_m T_d}|\psi_m\rangle$
\bea\label{epsilonm}
\epsilon_m= NJ\Big(1-\frac{\omega_d}{2n}\Big)+m\frac{\omega_d}{n},
\eea 
where $\omega_d={2\pi}/{T_d}$ is the driving frequency.  The set of $n$ Floquet eigenstates $|\psi_m\rangle$ $ (m=0,1,\cdots,n-1)$ form a multiplet in the quasienergy spectrum with equidistant quantized quasienergy $\omega_d/n$ as illustrated in Fig.~\ref{Fig-CMPinFTC}(left).

Note that the short-ranged correlated basis $|\phi_q\rangle$ in Eq.~(\ref{psim})  are not orthogonal for a finite-size system, i.e., 
$
\langle \phi_q  |\phi_{q'}\rangle=\big[\cos\frac{(q-q')\pi}{n}\big]^N.
$
However, in the thermodynamic limit $N\rightarrow \infty$, all the short-ranged correlated states $|\phi_q\rangle$ are orthogonal $\langle \phi_q  |\phi_{q'}\rangle=\lim_{N\rightarrow \infty}\big[\cos\frac{(q-q')\pi}{n}\big]^N\rightarrow \delta_{qq'}$. The long-range correlated eigenstates $|\psi_m\rangle$ are more like multi-component Schr\"odinger-cat states, which are unlike to survive in the thermodynamic limit. If the system is initially prepared in one Floquet eigenstate $|\psi_m\rangle$, any local perturbation in the environment can induce the system to collapse randomly into one of the short-range correlated basis $|\phi_q\rangle$. As consequence, the $\mathbb{Z}_n$ symmetry of $|\psi_m\rangle$ is spontaneously broken in the thermodynamic limit.
From Eq.~(\ref{psim}), we can express the short-range correlated states with Floquet eigenstates 
\bea\label{phiq}
|\phi_q\rangle=\frac{1}{\sqrt{n}}\sum^{n-1}_{m=0}e^{-i2mq\tau}|\psi_m\rangle, \ \ \ \mathrm{with} \ \ \ q=0,1,\cdots,n-1.
\eea 
Using Eq.~(\ref{epsilonm}), we have the stroboscopic time evolution of  $|\phi_q\rangle$ as following 
\bea\label{phiqt}
|\phi_q(pT_d)\rangle=\frac{1}{\sqrt{n}}e^{-iNJ(1-\frac{\omega_d}{2n})pT_d}\sum^{n-1}_{m=0}e^{-i2mq\frac{\pi}{n}-im\frac{\omega_d}{n}pT_d}|\psi_m(0)\rangle.
\eea 
We see that $\langle\phi_q(nT_d)|\hat{\Phi}|\phi_q(nT_d)\rangle=\langle\phi_q(0)|\hat{\Phi}|\phi_q(0)\rangle$ with integer $n>1$.  According to the definition Eq.~(\ref{psinT}), the DTTS is spontaneously broken in the thermodynamic limit accompanied by the breaking of a global internal $\mathbb{Z}_n$ symmetry described by $R_\tau$. The symmetry-broken localised states are also called {\it subharmonic modes} since their oscillating frequency is a fraction $\omega_d/n$ of the driving frequency $\omega_d$. The $\mathbb{Z}_n$ symmetry breaking (period-$n$ tupling) was discussed, e.g., in the $n$-hands clock model \cite{sreejith2016prb,surace2019prb}, ultracold bosons in a ring lattice \cite{pizzi2019prl} and long-range interacting systems \cite{pizzi2019arxiv}.

Since the original time-dependent Hamiltonian also satisfies $H(t)=H(t+nT_d)$, we have the freedom to choose the Floquet period $T=nT_d$. In this setting, the Floquet Hamiltonian is given formally by
\bea\label{HFnt}
H_F=i\frac{1}{nT_d}\ln\Big(\mathcal{T}\exp\big[-i\int_{0}^{nT_d}H(t)dt\big]\Big).
\eea
Under this choice,  the $n$ Floquet multiplets are ``degenerate" within the Floquet-Brillouin zone $0\leq\epsilon_\alpha < \omega_d/n$ as illustrated by the black lines in Fig.~\ref{Fig-CMPinFTC}(middle). They are long-range correlated eigenstates analogous to multi-component Schr\"odinger-cat states with $\mathbb{Z}_n$ symmetry as shown by Eq.~(\ref{psim}). In the thermodynamic limit, the $n$ long-range correlated states randomly collapse into one of the $n$ short-range correlated states, which are illustrated by the coloured wave packets in Fig.~\ref{Fig-CMPinFTC}(middle).  As consequence, the $\mathbb{Z}_n$ symmetry of long-range correlated eigenstates is spontaneously broken, and the resultant $n$ short-range correlated states exhibit oscillations with the same time period $nT$ but with phases differed by $2\pi/n$ [see Eqs.~(\ref{phiq}) and (\ref{phiqt})]. We arrange the $n$ short-range correlated states on a $\mathbb{Z}_n$ circle as shown in Fig.~\ref{Fig-CMPinFTC}(middle). The emergence of a multiplet of Floquet eigenstates with equidistant quasienergy $\omega_d/n$ as illustrated in Fig.~\ref{Fig-CMPinFTC}(left) is the prerequisite for DTTS breaking \cite{khemani2016prl,else2016prl,von2016prb}. 

\subsection{Stabilising mechanisms}

However, the subharmonic modes do not necessarily mean the Floquet time crystals. In the spin chain model discussed above, the subharmonic modes are generally unstable since the system is generically heated up to a featureless infinite-temperature state by the driving \cite{lazarides2014pre,dalessio2014prx,ponte2015aop}. As a solution, disorder [e.g., setting $W>0$ for random field $B^z_i$ in Eq.~(\ref{ht})] is used to stabilise the subharmonic modes \cite{khemani2016prl,else2016prl,von2016prb,yao2017prl} via the mechanism of many-body localization (MBL) \cite{nankishore2015arcmp,ponte2015prl,lazarides2015prl,abanin2016aop}.  To determine if the system is in the MBL phase, one can calculate the quasienergy statistics ratio, $\langle r \rangle=\min(\Delta_m,\Delta_{m+1})/\max(\Delta_m,\Delta_{m+1})$, where $\Delta_m=\epsilon_{m+1}-\epsilon_m$ is the $m$th quasienergy gap \cite{pal2010prb,khemani2016prl}, by averaging over both disorder and the quasienergy spectrum. In the thermal phase, this value approaches the circular orthogonal ensemble of $\langle r \rangle \approx 0.527$ while, in the MBL phase, the value should approach the Poisson limit of $\langle r \rangle \approx 0.386$ \cite{dalessio2014prx}. If the system is in the MBL phase, it was shown that \cite{yao2017prl} the spontaneous breaking of DTTS is rigid for a certain range of driving imperfection $0<\delta<\delta_c$. If the imperfection is larger than the critical point $\delta>\delta_c$, the system is still in the MBL phase but the DTTS is not broken. 
{
%

Floquet time crystals can also exist in clean Floquet systems without disorder \cite{Russomanno2017prb,huang2018prl,rovny2018prl,rovny2018prb,ho2017prl,barfknecht2019prb,pizzi2019arxiv}.  Although the fate of a generic driven many-body system is a trivial infinite temperature state with no long-range order, there exists a prethermal state with an extremely long lifetime provided that the driving frequency is much larger than the local energy scales in the system \cite{abanin2015prl,abanin2017prb,mori2016prl,kuwahara2016aop}. The system needs many local rearrangements of the degrees of freedom to absorb an ``energy quanta" from the driving, which is a high-order process taking exponentially long time even in a clean system with no MBL and without disorder. The DTTS broken state in this long time prethermal process is called {\it prethermal time crystal} \cite{else2017prx,luitz2019prethermalization}. 

By coupling the Floquet many-body system to a cold bath, it is also possible to protect the prethermal time crystals for infinitely-long time \cite{else2017prx}. Actually, the spontaneous breaking of DTTS was already discussed even before Wilczek's seminal work in a modulated dissipative cold atom system \cite{heo2010pre}, and the DTTS symmetry breaking has been reported in the experiment \cite{kim2006prl}. However, strictly speaking, Floquet time crystals by our definition are the quantum phenomena emerging in {\it closed} systems not in open systems. Therefore, we exclude the ``time order'' or the self-organised phenomena in driven-dissipative systems introduced by Prigogine \cite{Prigogine777}, such as synchronisation \cite{Kuramoto1987} and Belousov-Zhabotinsky reaction \cite{BZreaction}, from the family of Floquet time crystals.

In fact, MBL with disorder spin systems is NOT a necessary condition to realise Floquet time crystals in other physical systems (e.g., the quantum gas). 
There exist other mechanisms to break $\mathbb{Z}_n$ symmetry and stabilise the broken states other than MBL. Imagine the Floquet Hamiltonian (\ref{HFnt})  has a $\mathbb{Z}_n$-lattice potential along the $\mathbb{Z}_n$ circle as illustrated by the blue curve in Fig.~\ref{Fig-CMPinFTC}(right). If the lattice potential is strong enough, the Floquet eigenstates are the superposition of
localised states near the lattice sites (Wannier functions), which can tunnel to their neighbouring sites with hopping rate $J$.
In this case, the system can be described by a tight-binding model. As a result, the quasienergies of the Floquet eigenstates form a band ($ \propto |J|$) as indicated in Fig.~\ref{Fig-CMPinFTC}(right). 
The tunnelling rate can be tuned by system parameters.
In the limit $J\rightarrow 0$, the Floquet eigenstates become degenerate and form a multiplet ready for symmetry breaking. At this point, any nonequilibrium fluctuations can induce the Floquet eigenstates to collapse into one of the localised states, and the broken state is protected by the local potential well of $\mathbb{Z}_n$ lattice. 
If the subharmonic modes have attractive interaction, the spontaneous symmetry breaking occurs when the interaction strength is larger than a critical value proportional to the tunnelling rate $J$ \cite{sacha2015pra}. 
This mechanism of realizing time crystals without referring to MBL has also been studied in the driven Lipkin-Meshkov-Glick (LMG) model \cite{Russomanno2017prb}, where only the collective degree of freedom of spins is relevant due to conservation laws and in the thermodynamic limit the collective spins behave classically, therefore subharmonic response is protected by nonlinear resonances.  Time crystals can also be protected by Floquet dynamical symmetry (FDS) and robust against perturbations respecting the FDS symmetry \cite{chinzei2020arxiv}.


%







\section{Condensed matter physics in time crystals}

Most works so far on time crystals focus on the spontaneous breaking process of DTTS and the possible mechanisms to stabilise the subharmonic mode. However, the interplay of subharmonic modes is still not fully explored. We call the many-body physics of subharmonic modes the field of {\it condensed matter physics in time crystals}. In this section, we review the very preliminary results in this field based on the models of periodically driven quantum oscillators (e.g., driven quantum gas), instead of periodically driven spin models since there is no such work in spin models yet.
In general, the Hamiltonian of driven quantum oscillators in one dimension is described by
\begin{equation}\label{ManybodyH}
H(t)=\sum_{i}H_s(\hat x_i, \hat p_i,t)+\sum_{i<j}V(\hat x_i- \hat x_j),
\end{equation}
where $H_s({\hat x}_i,{\hat p}_i,t)$ is the single-particle Hamiltonian which is NOT assumed to be periodic in space but periodically time-dependent $H_s( \hat x_i,\hat p_i,t)=H_s( \hat x_i, \hat p_i,t+T_d)$ with $T_d=2\pi/\omega_d$ the driving period. The two-body interaction potential $V( \hat x_i- \hat x_j)$ depends on the distance of two particles. We first discuss the physics of single-particle Hamiltonian from Sec.~\ref{sec:psl} to Sec.~\ref{sec:top}, and then include the effects of interaction from Sec.~\ref{sec:i} to Sec.~\ref{sec:dl}.

\subsection{Phase space lattices}\label{sec:psl}

When talking about condensed matter physics, there is usually a lattice symmetry assumed in the Hamiltonian. However, Hamiltonian (\ref{ManybodyH}) has no such lattice structure in real space, and the time periodicity is also not the lattice structure we refer to since this time periodicity is already included in the Floquet theory. In fact, the lattice structure for the condensed matter physics in Floquet time crystals is hidden in phase space.
The general idea goes as follows \cite{sacha2018prl}. Let us begin with a classical single-particle Hamiltonian described by a time-independent $H_0(x,p)$. Since the energy is conserved in this case, the classical orbit is just the isoenergetic contour of $H_0(x,p)$. We define the {\it action} variable by 
\bea
I\equiv \frac{1}{2\pi}\oint p\,\mathrm{d}x
\eea
 along the isoenergetic contour of $H_0(x,p)$ and rewrite the Hamiltonian as a function of action $H_0(I)$. The conjugate variable of action $I$ is called {\it angle}  $\theta$, which does not explicitly appear in $H_0(I)$. According to the canonical equation of motion, we have $\theta(t)=\Omega t+\theta(0)$ with $\Omega(I)=dH_0(I)/dI$ representing the frequency of the regular motion orbit. 
 
 We further introduce a perturbation term of the form $H_1=\lambda h(x)f(t)$ where the time periodic function reads $f(t+2\pi/\omega_d)=f(t)=\sum_kf_ke^{ik\omega_d t}$ with $f_k=f^*_{-k}$ and the spatial function can be expressed in terms of the action and angle variables $h(x)=\sum_lh_l(I)e^{il\theta}$. Thus the whole single-particle Hamiltonian is given by,
\be\label{HItheta}
H_s(t)=H_0(I)+\lambda \sum_{l,k}h_l(I)f_k e^{il\theta+ik\omega_d t}.
\ee
Now if we tune the driving frequency to meet the resonant condition $\omega_d=n\Omega(I_0)$ with $n \in  \mathbb{N} $ and $I_0$ the action of a resonant orbit, in the vicinity of this resonant trajectory the motion of the particle can be separated into a fast oscillating and a slow varying parts, where the latter can be described by a new angle variable $\Theta=\theta-\omega_d t/n$. Using the generating function $G_2=I(\theta-\omega t/n)$, we transform into the rotating frame with frequency $\omega_d/n$ and retain the time-independent Hamiltonian from Eq.~(\ref{HItheta}) in the rotating wave approximation (RWA)
\bea
H_{RWA}(I,\Theta)&=&H_0(I)-\frac{\omega_d}{n}I+\lambda\sum_kh_{kn}(I)f_{-k}e^{ikn\Theta} \nonumber\\
&\equiv& H_0(I)-\frac{\omega_d}{n}I+\lambda V_{\mathrm{eff}}(I,\Theta)
\eea
where $V_{\mathrm{eff}}(I,\Theta)=\sum_kh_{kn}(I)f_{-k}e^{ikn\Theta}$ is a periodic potential in the angel direction $\Theta$ of rotating frame. This effective Hamiltonian essentially describes a particle moving on the $\Theta$ ring in the presence of a time-independent effective lattice potential $V_{\mathrm{eff}}(I,\Theta)$ which can be engineered by tuning the Fourier components $f_k$ and $h_l$ in the experiments. For example, for the monochromatic driving with frequency $\omega_d$, the effective lattice potential is $V_{\mathrm{eff}}(I,\Theta)\propto\cos(n\Theta)$ meaning the driving resonates with the classical orbit with frequency $\omega_d/n$ and generates $n$ identical local wells uniformly distributed on the $\Theta$ ring. Below, we will introduce several concrete examples to show various lattice structures in phase space.

\begin{figure}
\centerline{\includegraphics[width=\linewidth]{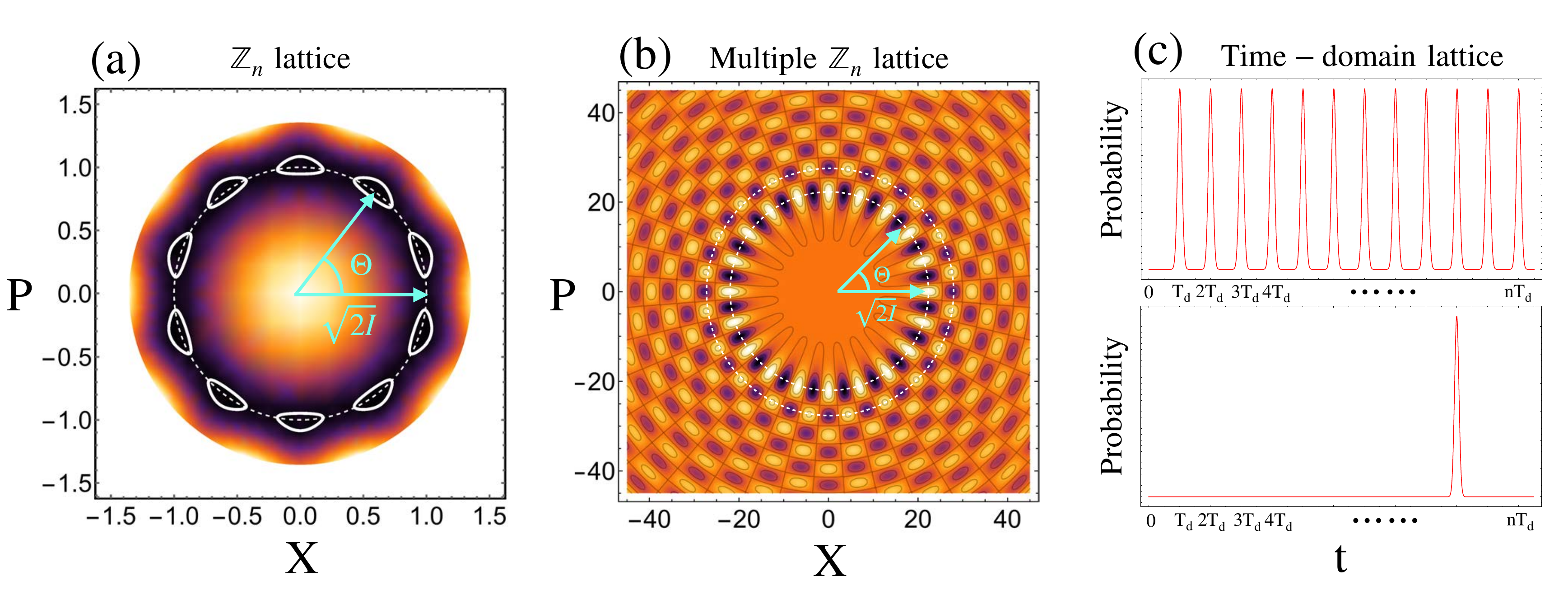}}
\caption{\label{Fig-PSLandTC}{\bf  Lattices in phase space and time domain.} (a) 2D density plot of Hamiltonian (\ref{HZn}) in phase space
for power $n=10$ and driving strength $\mu=0.3\mu_c$ with cut through the bottom of the quasienergy. The resultant $\mathbb{Z}_n$ lattice has $n$ stable states (white closed curves)
arranged periodically in angular direction $\Theta$, which is invariant under discrete phase space
rotations $e^{i\Theta} \rightarrow e^{i(\Theta+\tau)}$ where
$\tau=2\pi/n$. (b) 2D plot of Hamiltonian (\ref{g}) composed of many loops (white dashed circles) representing multiple $\mathbb{Z}_n$ lattices divided by the zero lines. There are $2n$ stable states
arranged periodically in angular $\Theta$ direction but the lattice constant is still $\tau=2\pi/n$, which means the whole lattice has two sublattices. (c) Probability to detect the system versus time at a fixed spatial point in the laboratory frame for the exact Floquet eigenstate (upper) and the subharmonic mode (lower). The exact Floquet eigenstate has the time period of the driving field ($T_d=2\pi/\omega_d$), but the subharmonic mode has a longer time period of $nT_d$ breaking the DTTS.   }
\end{figure}

The first example is a nonlinear oscillator driven
by an external field coupling nonlinearly to the coordinate, with the single-particle
Hamiltonian reading \cite{guo2013prl}
\begin{equation}\label{DDO}
H_s(t)=\frac{{p}^2}{2m}+\frac{1}{2}m\omega_0^2 {x}^2 + \frac{\nu}{2}
{x}^4+2f\cos(\omega_d t){x}^n.
\end{equation}
Here $\omega_0$ is  the frequency of the oscillator and $\omega_d$
the driving frequency. The nonlinearity of coupling is characterized by the
exponent $n$.  If $n=1$ the model (\ref{DDO}) is the linearly
driven Duffing oscillator \cite{dykman1975zetf,peano2006cp,guo2010epl,guo2011pre}, for $n=2$  it is a
parametrically driven oscillator \cite{dykman1998pre,marthaler2006pra,marthaler2007pra}. The model (\ref{DDO}) extends the exponent $n$ to be any integer larger than two, and has been recently realised up to $n=5$ in the experiment with superconducting circuits \cite{svensson2018apl}.
We are interested in the subresonance regime, where the driving frequency $\omega_d$ is close to $n$
times $\omega_0$, i.e., the detuning 
$\delta\omega\equiv\omega_0-\omega_d/n$ is much smaller than
$\omega_0$. By taking the Harmonic oscillator as $H_0(x,p)$, the time-independent RWA Hamiltonian with action and angle variables is given by 
\bea\label{HZn}
H_{RWA}(I,\Theta)=\big(I-\frac{1}{2}\big)^2+\mu(\sqrt{2I})^n\cos(n\Theta).
\eea
Here, the Hamiltonian has been scaled by energy unit $4m^2\omega^2_0\delta\omega^2/3\nu$ and $\mu$ is the scaled driving strength given by $\mu=\frac{3\nu f}{2}\left( \frac{m\omega_0\delta
\omega}{-3\nu}\right)^{(n-4)/2}$ with red detuning $\delta\omega<0$. The RWA Hamiltonian (\ref{HZn}) displays a new symmetry not visible in
(\ref{DDO}). To show the lattice structure, we define two alternative conjugate variables 
\bea
X\equiv \sqrt{2I}\cos\Theta, \ \ \ P\equiv \sqrt{2I}\sin\Theta
\eea
and plot $H_{RWA}$ in phase space spanned by $X$ and $P$ in Fig.~\ref{Fig-PSLandTC}(a). Clearly, there are $n$ identical local wells distributed uniformly on a circle with radius $\sqrt{2I_0}$ with $I_0=1/2$. The Hamiltonian shows a $\mathbb{Z}_n$ lattice structure in phase space. The $n$ local wells only exist in a certain range of the driving strength $0<|\mu|<\mu_c$, where the critical driving strength is given by $\mu_c=(1-r_c^2)/(nr^{n-2}_c)$ with $r^2_c=(n-2)/(n-4)$ \cite{guo2013prl}.

The second example is a harmonic oscillator in driven optical lattice, where the power-law driving in model (\ref{DDO}) is replaced by a cosine-type driving,
i.e., the single-particle Hamiltonian is
given by  \cite{guo2016njp}
\begin{equation}\label{DrivenOpticalLattice}
H_s(t)=\frac{{p}^2}{2m}+\frac{1}{2}m\omega_0^2{x}^2+2A\cos(k{x}+\omega_dt).
\end{equation}
The above Hamiltonian also describes a cavity mode driven by dc-biased Josephson junction(s) \cite{hofheinz2011prl,armour2013prl,gramich2013prl}, 
which has triggered a new research field of Josephson photonics \cite{cassidy2017science,armour2017prb,westig2017prl,rolland2019prl}.
Under the subresonance condition $\omega_d=n\omega$, the RWA Hamiltonian in the classical limit with action and angle variables is given by \cite{guo2016njp}
\begin{equation}\label{g}
H_{RWA}(I,\Theta)=2\mu J_n(\sqrt{2I})\cos(n\Theta-\frac{n\pi}{2}).
\end{equation}
where $J_n(\bullet)$ is
the Bessel function of order $n$. The Hamiltonian has been scaled by energy unit $m\omega^2/k^2$ and $\mu\equiv Ak^2/m\omega^2$ is the dimensionless driving strength. 
In Fig.~\ref{Fig-PSLandTC}(b), we plot the scaled RWA Hamiltonian in phase space. The whole lattice
structure clearly shows a $\mathbb{Z}_n$ symmetry. The zero
lines (i.e., $H_{RWA}=0$) divide the whole
phase space lattice into many small `` cells ". The center of each
cell is a stable point corresponding to either a local minimum or
a local maximum. The area inside the cell represents the basin of attraction
for the stable state in the center.
The RWA Hamiltonian shows a multiple $\mathbb{Z}_n$ lattice structure in phase space.

\begin{figure}
\centering
\includegraphics[width=\linewidth]{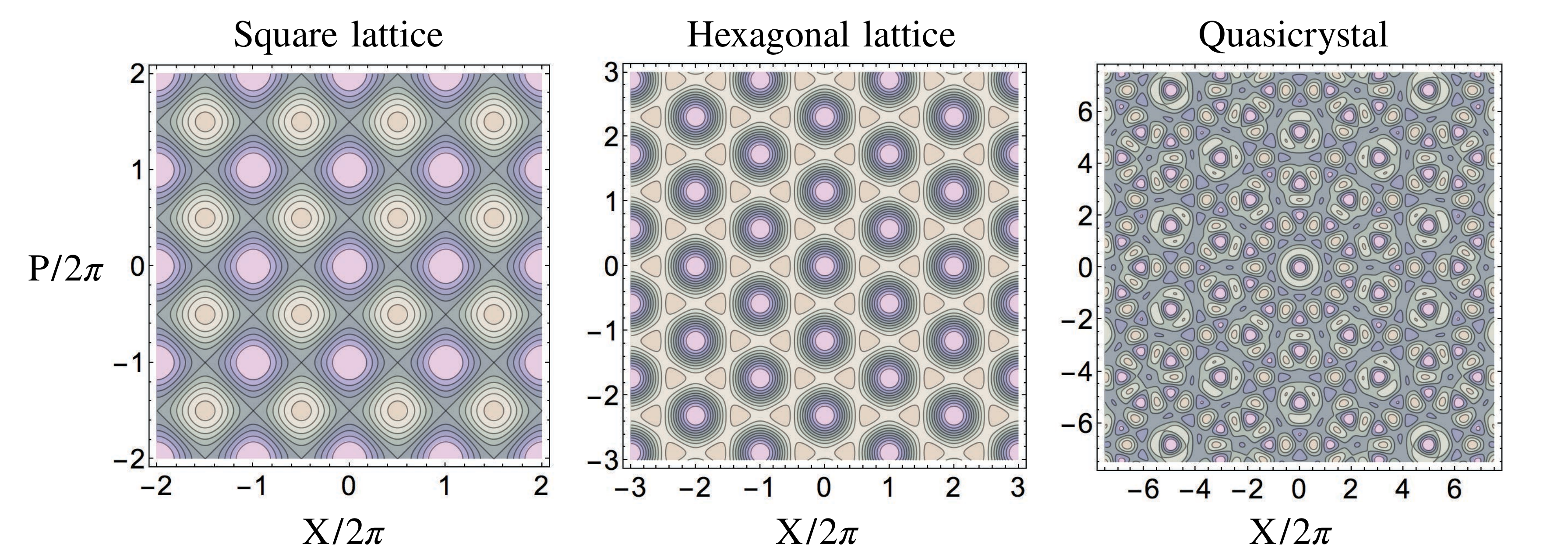}
\caption{\label{Fig-PSL-Crystal}{\bf Phase space lattices:} 2D density plot of Hamiltonian (\ref{eq:RWAHamiltonian}) showing
(a) square lattice for $q_0=4$, (b) hexagonal lattice for $q_0=3$ or $q_0=6$ and (c) quasicrystal structures for $q_0=5$. The value of $H_{RWA}(X,P)$ has been scaled by the kicking strength $K$ in all figures.}
\end{figure}

In the above two examples, the periodic structure is actually one dimensional, i.e., the lattice structure appears along the angular $\Theta$ direction in phase space.  One may ask if it is possible to create a two dimensional lattice in the whole phase space. The answer is yes \cite{liang2018njp}. Let us consider the famous model of kicked harmonic oscillators (KHO) \cite{zaslavsky1986zetf,zaslavsky2008book,berman1991,buchleitner2004prl} with the single-particle Hamiltonian given by   
\begin{equation}\label{eq:dimensionlessKHO}
H_s(t)=\frac12({x}^2+{p}^2)+K\tau\cos x\sum_{n=-\infty}^{\infty}\,\delta(t-n\tau),
\end{equation}
where $K$ is the dimensionless kicking strength and $\tau$ is the dimensionless kicking period.
In the resonant condition that the kicking period satisfies $\tau=2\pi/q_0$ with $q_0$ an integer and the weak kicking strength limit$|K|\ll 1$, the single-particle dynamics is still dominated by the fast harmonic oscillation but with slowly modulated amplitude and phase. 
The RWA part of the single-particle Hamiltonian is given by \cite{liang2018njp}
\begin{equation}\label{eq:RWAHamiltonian}
H_{\it RWA}(X,P)=\frac{K}{q_0}\sum_{j=1}^{q_0}\,\cos\left( X\cos\frac{2\pi j}{q_0}+ P\sin\frac{2\pi j}{q_0}\right).
\end{equation}
In Fig.~\ref{Fig-PSL-Crystal}, we plot $H_{\it RWA}(X,P)$ in phase space for different $q_0$. We see that the $H_{\it RWA}(X,P)$ has a square lattice structure for $q_0=4$, hexagonal lattice structure for $q_0=3$ or $q_0=6$, and even quasicrystal structure for $q_0=5$ or $q_0\geq7$. 

There is another way to view the hidden lattice structure of Hamiltonian. In the laboratory frame, the crystalline structure in $\Theta$ is reproduced in the time domain as the relation between $\Theta$ and $t$ is linear,  i.e., $\Theta=\theta-\omega_d t/n$. Imagine we measure the probability of the system by locating a detector at a fixed spatial point close to the classical resonant trajectory, e.g., at point $(1,0)$ in Fig.~\ref{Fig-PSLandTC}(a). If the system is in the subharmonic mode, i.e., a localised wave packet at one site of the $\mathbb{Z}_n$ lattice, the detector will be triggered with longer time period $nT_d$ breaking the original DTTS of driving field as indicated by the lower subfigure in Fig.~\ref{Fig-PSLandTC}(c). If the system is in one Floquet eigenstate, which is the superposition of $n$ localised states in the $\mathbb{Z}_n$ lattice, the detector will respond with the driving period $T_d$ as indicated by the upper subfigure in Fig.~\ref{Fig-PSLandTC}(c).
This time-domain lattice picture was introduced by Sacha \cite{sacha2015scirep} to study cold atoms bouncing on an oscillating mirror. In this picture, the roles of space and time are exchanged making it possible to discuss analogous condensed matter phenomena like Anderson localisation and Mott insulator in the time domain. Here in this review, we keep our discussion in the rotating frame and adopt the picture of phase space lattices. In quantum dynamics, one has to find a unitary operator to describe the discrete lattice transformation under which the RWA Hamiltonian is invariant. Such operator is natural to be defined in the picture of phase space lattice, and allows us to develop a band theory, which is to be discussed in the next section. 

\begin{figure}
\centerline{\includegraphics[width=\linewidth]{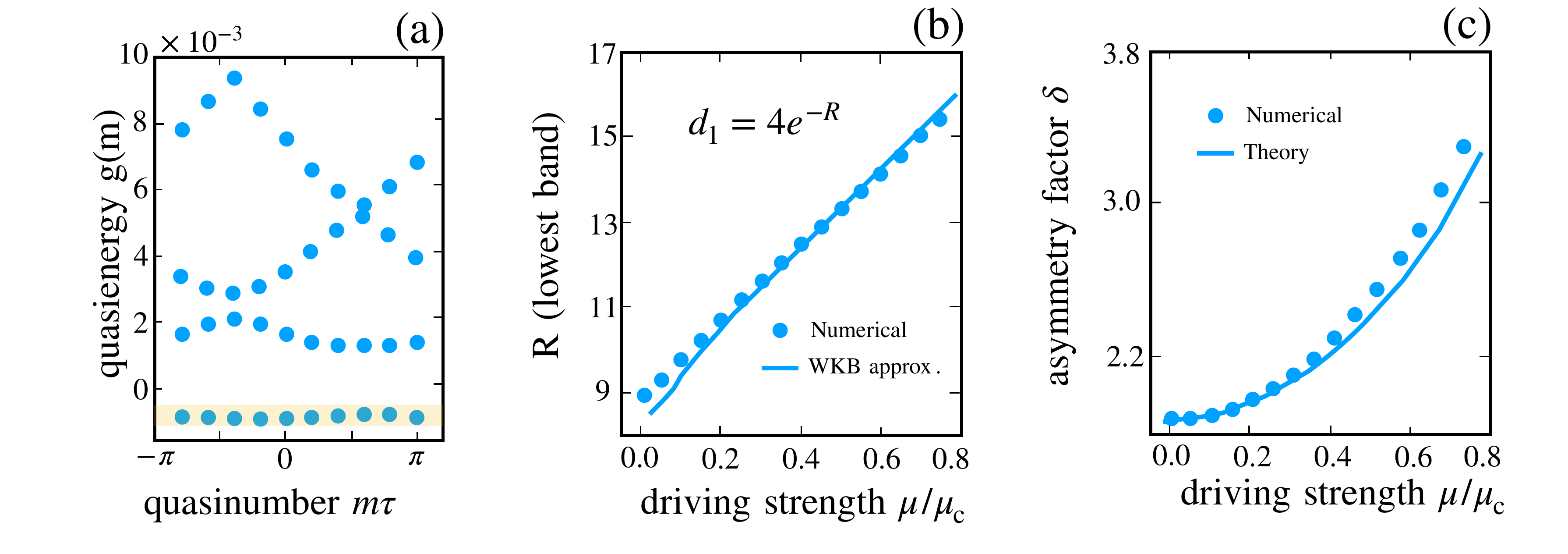}}
\caption{\label{Fig-ZnLattice-Bands}{\bf Quasienergy band structure.} (a) Quasienergy band structure in the reduced Brillouin zone of quasinumber $m$ with the lowest band marked in yellow.
(b) Bandwidth of the lowest band $d_1=4e^{-R}$ versus driving strength from numerical simulation (solid dots) and WKB calculation (blue lines). 
(c) Asymmetry factor $\delta$ of the lowest band versus driving strength from numerical simulation (solid dots) and theory (blue lines). 
Parameters: $\lambda=1/205$,
$n=10$ for all the figures,
$\mu=0.3\mu_c$ for figure (a). 
}
\end{figure}

\subsection{Band theory}\label{sec:bt}

In the above section, we discuss the phase space lattices in the classical limit. In quantum mechanics, the lattice symmetry results in the band structure of {\color{blue}the} energy spectrum. Let us start from the example Hamiltonian (\ref{DDO}).
We transform into the rotating frame via the unitary operator $\hat{U}=e^{i({\omega_d}/{n}) \hat{a}^\dagger \hat{a}
t}$, where $\hat{a}$ is the annihilation operator of the
oscillator, resulting in
$$
{H}_{RF}(t)
=\hat{U}{H}(t)\hat{U}^\dagger+i\hbar\dot{\hat{U}}\hat{U}^\dagger={H}_{RWA}+{H}_{non-RWA}.
$$
The
RWA part Hamiltonian is given by (scaled by energy unit of $4m^2\omega^2_0\delta\omega^2/3\nu$)
\begin{equation}\label{H}
{H}_{RWA}=-\lambda {\hat{a}^\dagger}
\hat{a}+\lambda^2{\hat{a}^\dagger}
\hat{a} ({\hat{a}^\dagger} \hat{a}+1)+
\frac{1}{2}\mu(2\lambda)^{\frac{n}{2}}\Big({\hat{a}^{\dagger
n}}+\hat{a}^n\Big). 
\end{equation}
Here,  $\lambda\equiv-3\nu\hbar/(4m^2\omega_0^2\delta\omega)$ is the scaled dimensionless Planck constant, which describes the quantumness of our system. In the classical limit $\lambda\rightarrow 0$, expression (\ref{H}) goes back to the classical form (\ref{HZn}). The non-RWA part has the following explicit form \cite{guo2013prl}
\begin{eqnarray}\label{hnonrwa}
{H}_{non-RWA}&=&\frac{1}{3}
\lambda^2(2{\hat{a}^\dagger}\hat{a}-1)\hat{a}^{\dagger2}e^{i\frac{2\omega_dt}{n}}+\frac{1}{6}\lambda^2\hat{a}^{\dagger4}e^{i\frac{4\omega_dt}{n}}\nonumber\\
&&+
\frac{1}{2}\mu(2\lambda)^{\frac{n}{2}}\Big[\Big({\hat{a}^{\dagger
}e^{i\frac{2\omega_dt}{n}}}+\hat{a}\Big)^n-\hat{a}^n\Big]+\emph{h.c.}.
\end{eqnarray}
According to the Floquet-Magnus expansion (\ref{h0h1h3}), the RWA Hamiltonian
$\hat{H}_{RWA}$ is just the leading order of Floquet Hamiltonian, i.e., $\hat{H}_{RWA}=H^{(0)}_F=T^{-1}\int_0^T \hat{H}_{RF}(t)\mathrm{d}t$ with Floquet period $T=n2\pi/\omega_d$.

The RWA Hamiltonian (\ref{H}) displays a new symmetry not visible in the original Hamiltonian
(\ref{DDO}). We define a unitary operator
$
\hat{T}_\tau\equiv e^{-i\tau {\hat{a}^\dagger} \hat{a}}, 
$
which has the
properties $\hat{T}_\tau^\dagger a \hat{T}_\tau = a e^{-i\tau}$
and $\hat{T}_\tau^\dagger a^n \hat{T}_\tau = a^n e^{-in\tau}$. It
is direct to find that the RWA Hamiltonian is invariant under
discrete transformation 
\bea\label{THT}
T_\tau^\dagger \hat{H}_{RWA} T_\tau =
\hat{H}_{RWA},\ \  \mathrm{for} \ \ \tau={2\pi}/{n}.
\eea
According to the Bloch's theorem, the eigenstates
$\psi_m(\Theta)$ of the RWA Hamiltonian,
$\hat{H}_{RWA}\psi_m(\Theta)=g(m)\psi_m(\Theta)$,
 have the form $
\psi_m(\Theta)=\varphi_m(\Theta)e^{-im\Theta}$, with a periodic
function $\varphi_m(\Theta+\tau)=\varphi_m(\Theta)$. Here, the
integer number $m$, which is called \textit{quasi-number} in Ref.~\cite{guo2013prl}, plays the
role of the quasi-momentum
$k$ in a crystal.
While the quasi-momentum
$k$ is conjugate to
the coordinate, the quasi-number $m$ is conjugate to the phase
$\Theta$.  

In Fig.~\ref{Fig-ZnLattice-Bands}(a), we plot the quasienergy band structure in
the reduced Brillouin zone $m\tau\in(-\pi,\pi]$ from exact numerical diagonalization. Here we relabel
the eigenstates $\psi_m(\Theta)$ by $\psi_{lm}(\Theta)$, where
$l=1,2,...$ is the label of the bands counted from the bottom.
For finite values of $n$ (in our numerical simulation we chose $n=10$)
the quasienergy band spectrum is discrete. It would become more continuous
in the limit of large $n$.
For the multiple $\mathbb{Z}_n$ lattice Hamiltonian (\ref{DrivenOpticalLattice}), a similar quasi-number band theory was developed in Ref.~\cite{guo2016njp}.


\begin{figure}
  \centering
  \includegraphics[width=0.6\linewidth]{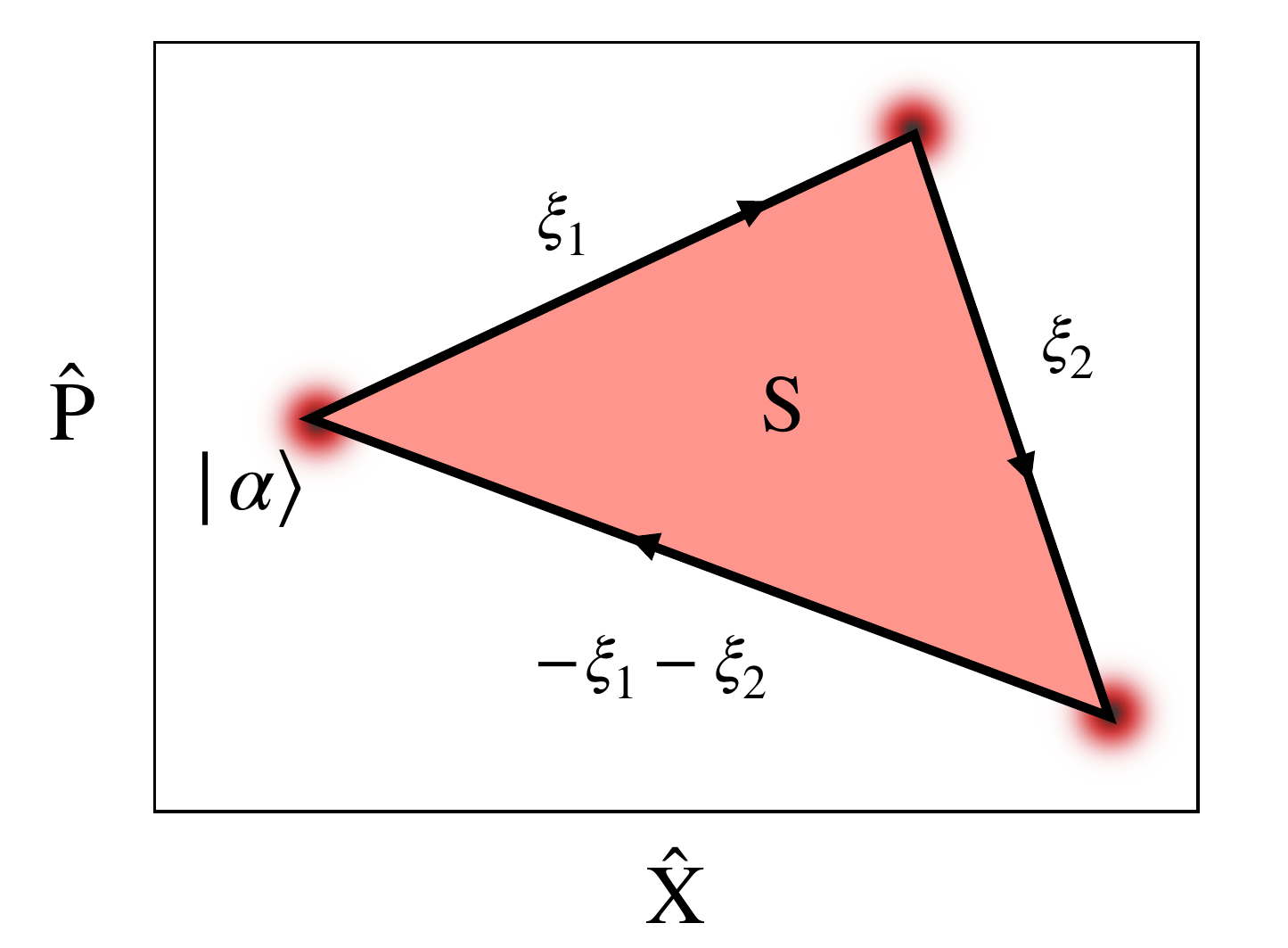}
 \caption{\label{fig_QunatumPhase}{\bf Geometric phase in noncommutative phase space.} A coherent state $|\alpha\rangle$ is moved along a closed triangle by three displacement operators, i.e., $D[-(\xi_1+\xi_2)]D(\xi_2)D(\xi_1)|\alpha\rangle=e^{i\frac{1}{\lambda}S}|\alpha \rangle,$ where $D(\xi)\equiv\exp(\frac{\xi}{\sqrt{2\lambda}} \hat{a}^\dagger-\frac{\xi^*}{\sqrt{2\lambda}}\hat{a})$ with two complex numbers $\xi_1, \ \xi_2$ determining the moving path. The geometric phase factor is given by $e^{i\frac{1}{\lambda}S}$ with $S=\frac{1}{2}\mathrm{Im}[\xi_2\xi^*_1]$ being the area of the enclosed path (red area). }
\end{figure}

\subsection{Tight-binding model}\label{sec:tbm}

The phase space lattice is fundamentally different from the real space lattice as the two degrees of freedom of phase space do not
 commute, i.e., $[\hat X,\hat P]=i\lambda$ for the Hamiltonian (\ref{H}). As a result, the band structure of a phase space lattice shows some novel features. 
For the band structure shown in Fig.~\ref{Fig-ZnLattice-Bands}(a), the spectrum of the lower $l$-th band is approximately given by a tight-binding model \cite{guo2013prl}
\begin{eqnarray}\label{spectruminoneband}
g_l(m)=E_l-2|J_l|\cos(m\tau+\delta\tau),
\end{eqnarray}
centered around  $E_l$ and with bandwidth $d_l=4|J_l|$. The result
shows a surprising asymmetry. From the plot of the phase space lattice in
Fig.~\ref{Fig-PSLandTC}(a) we would have expected a degeneracy $g(m)=g(-m)$, since
clockwise and anticlockwise motion should be equivalent, as in the
case of orbital motion. However, in the noncommutative phase space, the concept of point is meaningless. Instead, we should define a coherent state $|\alpha\rangle$ which is the eigenstate of the lowering operator, i.e., $\hat{a}|\alpha\rangle=\alpha|\alpha\rangle$ with $\hat{a}\equiv(\hat{X}+i\hat{P})/\sqrt{2\lambda}$. As shown in Fig.~\ref{fig_QunatumPhase}, we observe that a coherent state moving along a closed path in phase space naturally acquires an additional quantum phase factor  $e^{iS/\lambda}$, where $S$ is the enclosed area \cite{pechal2012prl}. This geometric phase is responsible for the asymmetry of quasienergy band structure, and also called Peierls phase \cite{hofstadter1976prb} for describing the hopping of charged particles between tight-binding lattice sites in a magnetic field.  For neutral ultracold atoms in an optical lattice, a controlled Peierls phase can be created by shaking the lattice to realize artificial gauge fields \cite{dalibard2011rmp,struck2012prl,goldman2014rpp}.

According to the Bloch theorem, the tight-binding eigenstate is 
\bea
|\psi_{lm}\rangle=\frac{1}{\sqrt{n}}\sum_{q=0}^{n-1}e^{imq\tau}{T_\tau^q}|\phi_{0,l}\rangle
\eea
where $|\phi_{q,l}\rangle$ is the $l$-th localised Wannier level in the $q$-th well of the $\mathbb{Z}_n$ lattice, and $\hat{T}_\tau=e^{-i\tau {\hat{a}^\dagger} \hat{a}}$ is the discrete rotation operator shown in Eq.~(\ref{THT}).
Reminiscent of eigenstate expression (\ref{psim}) for the spin model, they both reflect the underlying $\mathbb{Z}_n$ symmetry of corresponding systems. The nearest coupling in Eq.~(\ref{spectruminoneband}) is given by
\bea
J_l=-\int^{2\pi}_0[{T}_\tau\phi_{0,l}(\Theta)]^*{H}_{RWA}\phi_{0,l}(\Theta) \mathrm{d}\Theta,
\eea
where we have written the Wannier functions in the angle $\Theta$ representation. Interestingly, the calculation in Ref.~\cite{guo2013prl} shows the nearest neighbor coupling rate is in general a complex number
$J_l=|J_l|e^{i\tau\bar{r}^2/2\lambda}$ where $\bar{r}$ is the average radius. Hence the asymmetry factor is $\delta=\bar{r}^2/2\lambda \
(\mathrm{mod}\ n)$. In Fig.~\ref{Fig-ZnLattice-Bands}(c), we show the dependence of the asymmetry
factor $\delta$ on the driving strength $\mu$, obtained both in the
tight-binding calculation described above and from a numerical simulation. The asymmetry factor $\delta=\bar{r}^2/2\lambda \
(\mathrm{mod}\ n)$ in Eq.~(\ref{spectruminoneband}) from the imaginary part of $J_l$ is valid for large exponent $n\gg 1$. More precise calculation for small exponent (period tripling $n=3$) was given in Ref.~\cite{zhang2017pra}. The complex tunnelling parameter $J_l$ naturally arises in the plane of the phase space due to the noncommutative geometry.

The amplitude of tunnelling rate $J_l$  was calculated using the WKB semiclassical approximation in Ref.~\cite{guo2013prl}, i.e.,
$
 |J_l|=\frac{\lambda \omega_e}{2\pi}\exp\big(-\frac{r_s}{\lambda}W_l\big)
$
with the exponent
$W_l$
depending on the system parameters. Here, $r_s$ is the radius of the saddle points of $H_{RWA}$ and $\omega_e$ is the harmonic frequency near the local minimal points.
In Fig.~\ref{Fig-ZnLattice-Bands}(b) we compare
our approximate result for the first bandwidth $d_1=4|J_1|$ versus
driving $\mu$ to  numerical results. In the tight binding regime
they agree well with each other.
The tunnelling rate $|J_l|$ decreases exponentially to zero in the classical limit of $\lambda\rightarrow 0$. This means the $n$ eigenstates in each quasienergy band become degenerate and form the multiplet in the Floquet spectrum as shown in Fig.~\ref{Fig-CMPinFTC}(middle). In this limit, the Floquet eigenstate with $\mathbb{Z}_n$ symmetry cannot survive and will be broken into one of the localised Wannier states. Therefore, $\mathbb{Z}_n$ symmetry is spontaneously broken and the broken state oscillates with time period of $nT$, i.e., the subharmonic mode of Floquet time crystal. This period-multiplication phenomena predicted by the model (\ref{DDO}) has been verified in the experiment with superconducting circuits \cite{svensson2017prb,svensson2018apl}, where experimentalists observed robust output radiation (subharmonic oscillations \cite{wuatman2019ltp}) at a frequency close to a fraction $\omega_d/n$ of the driving frequency and evenly $2\pi/n$-shifted phase components. These observations call for further exploration of the quantum properties of the period-multiplication and a search for engineering multicomponent quantum superpositions of macroscopic coherent states (multicomponent cat-states). While most of these experimental findings display classical fixed points and the hopping between them induced by classical noise,  the quantum mechanical coherences in the form of multi-component cat-states are still absent. The $\mathbb{Z}_2$ symmetry breaking of parametrically driven model has been reported in the experiment with cold atoms \cite{kim2006prl} and analysed in Ref.~\cite{heo2010pre}. Recently, there is a lot of interest in $\mathbb{Z}_3$-lattice of triply driven oscillator \cite{svensson2017prb,zhang2017pra,zhang2019pre,lorch2019prr,gosner2019relaxation}, e.g.,  nonlocal random walk \cite{zhang2019pre},  quantum state preparation \cite{lorch2019prr} and dissipative phase transition \cite{gosner2019relaxation}.
A similar calculation for the band structure of the multiple $\mathbb{Z}_n$ lattice (\ref{g}) with tight-binding model has been provided in Ref.~\cite{guo2016njp}. There, the situation is more complicated since the tunnelling between nearest $\mathbb{Z}_n$ lattices also play an important role when bands cross each other. 

In the above discussion, we work in the single-particle picture. For an ensemble of identical particles free of interaction, we can describe the corresponding many-body Hamiltonian with the following tight-bing model
\bea\label{TBM}
H_{TB}\approx E_l\sum^{n-1}_{q=0}\hat a^\dagger_{q,l}\hat a_{q,l}+J_l\sum^{n-1}_{q=0}\hat a^\dagger_{q,l}\hat a_{q+1,l}.
\eea 
Here, $\hat a_{q,l}$ is the annihilation operator of the Wannier state $|\phi_{q,l}\rangle$.  We have restricted to the Fock space $|n_{0,l},n_{1,l},\cdots,n_{q=n-1,l}\rangle$ with $n_{q,l}$ the particle number occupying the Wannier state $|\phi_{q,l}\rangle$. 
A similar tight-binding model was introduced independently by K. Sacha in the study of bouncing atoms on an oscillating mirror with the subresonant codition \cite{sacha2015scirep}. 

\begin{figure*}
\centering
\includegraphics[width=\linewidth]{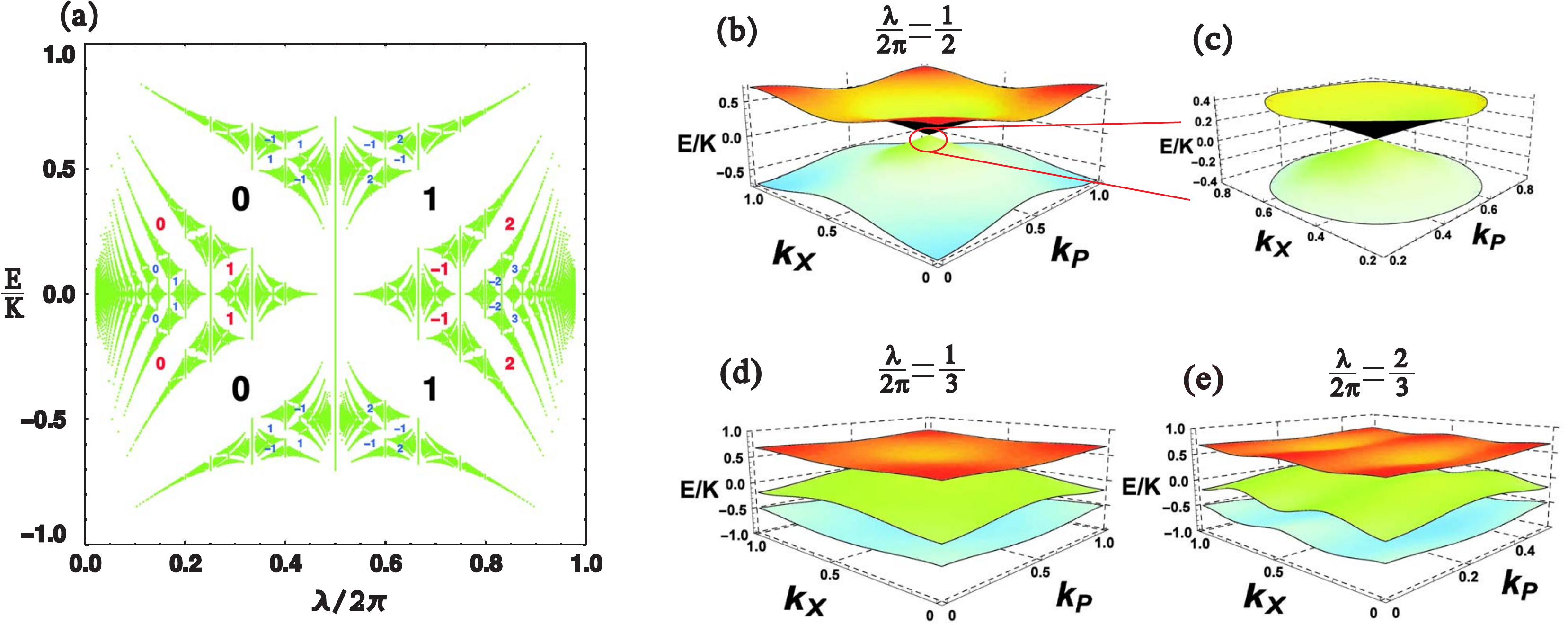}
\caption{\label{fig:bands}{\bf Topological band structures.} (a) Hofstadter's butterfly: the quasienergy spectrum of Hamiltonian (\ref{eq:Hamiltoniansq}) with rational number $\lambda/2\pi\in[0,1]$. The integers indicate the Chern number of each gap defined as the sum of Chern numbers of all the quasienergy bands below (or above) the gap. (b)-(e) Quasienergy band structures in 2D Brillouin zone for different parameters $\lambda$ which are given above the figures. For $\lambda/2\pi=1/2$, the two bands touch each other in the center of Brillouin zone. The linear dispersion relationship near the touching point is shown in figure (c).  The images were taken from Ref.~\cite{liang2018njp}.
}
\end{figure*}

\subsection{Topology}\label{sec:top}

Topological phenomena in condensed matter physics have been intensively investigated in the past decades. In quantum Hall physics, the origin of topology comes from the geometric phase induced by the applied magnetic field and the resultant band can be characterized by its topological invariant (Chern number or TKNN invariant), which relates to the quantized Hall conductance directly. In the noncommutative phase space, a geometric phase factor of a quantum state appears naturally when it moves along an enclosed path as shown in Fig.~\ref{fig_QunatumPhase}. We would expect that similar topological phenomena also exist in phase space lattices. This is actually true for the example Hamiltonian (\ref{eq:RWAHamiltonian}).  In the case of square lattice ($q_0=4$), Hamiltonian (\ref{eq:RWAHamiltonian}) is further simplified as
\begin{eqnarray}\label{eq:Hamiltoniansq}
H_{sq}(\hat X,\hat P)=\frac 12K\Big(\cos\hat X+\cos\hat P\Big).
\end{eqnarray}
This Hamiltonian is closely related to the established Harper's equation, which is a tight-binding model governing the motion of noninteracting electrons in the presence of a two-dimensional periodic potential and a uniformly threading magnetic field \cite{harper1955,hofstadter1976prb}.
The $H_{sq}(\hat X,\hat P)$ is invariant under discrete translation in phase space by two operators
$\hat{T}_1\equiv e^{i\frac{2\pi}{\lambda}\hat{P}}$ and $\hat{T}_2\equiv e^{i\frac{2\pi}{\lambda}\hat{X}}$.
However, the translation operators $\hat{T}_1$ and $\hat{T}_2$ are NOT commutative in general, i.e., $$[\hat{T}^r_1,\hat{T}^s_2]=\hat{T}_1\hat{T}_2(1-e^{-i4\pi^2rs/\lambda})$$ with integer powers $r,s\in\mathbb{Z}$ except for $2\pi rs/\lambda\in \mathbb{Z}$. If $\lambda/2\pi=p/q$ with $p$ and $q$ two coprime integers, we choose the following two commutative translational operators ($r=1$, $s=p$):
$
\hat{T}_X\equiv\hat{T}_1=e^{i\frac{2\pi }{\lambda}\hat{P}}, \   \hat{T}_P\equiv\hat{T}^p_2=e^{i\frac{2\pi p}{\lambda}\hat{X}}.
$
Therefore, we can search for the common eigenstates of the commutative operators $\hat{T}_X$ and $\hat{T}_P$ with eigenvalues given by $e^{i2\pi k_X}$ and $e^{i2\pi p k_P}$ respectively. The boundaries of the two dimensional Brillouin zone are defined by $0\leq k_X\leq1$ and $0\leq k_P\leq1/p$,
where $k_X$ and $k_P$ are quasimomentum and quasicoordinate, respectively \cite{zak1972ssp}. 

In Fig.~\ref{fig:bands}(a), we plot the quasienergy spectrum for $\lambda/2\pi\in[0,1]$, showing a Hofstadter's butterfly structure identical to that in quantum Hall systems. In Fig.~\ref{fig:bands}(b), (d) and (e), we plot the quasienergy band structures in the two-dimensional Brillouin zone $(k_X,k_p)$ for $\lambda/2\pi=1/2, 1/3$ and $2/3$, respectively. We see that the quasienergy band structure is two-fold degenerate for $\lambda/2\pi=2/3$ while there is no degeneracy for $\lambda/2\pi=1/2$ and $\lambda/2\pi=1/3$. In fact, for each rational $\lambda/2\pi=p/q$ (remembering $p,q$ are coprime integers), the spectrum contains $q$ bands and each band has a $p$-fold degeneracy \cite{dana1995pre}.
To show the underlying topology of the quasienergy bands, one can calculate the Chern number for a given band \cite{reynoso2017pra}
$$
c_b=\oint_\mathcal{C}\langle\psi_{b,\mathbf{k}}|\partial_\mathbf{k}|\psi_{b,\mathbf{k}}\rangle\cdot \mathrm{d}\mathbf{k},
$$
where the contour $\mathcal{C}$ is integrated over the boundary of the Brillouin zone. Accordingly, the Chern number of a gap below (above) the zero energy line can be defined as the sum of Chern numbers of all the quasienergy bands below (above) the gap \cite{liang2018njp}. The Chern number of some gaps are calculated and labelled in Fig.~\ref{fig:bands}(a).

\begin{figure*}
\centering
\includegraphics[width=\linewidth]{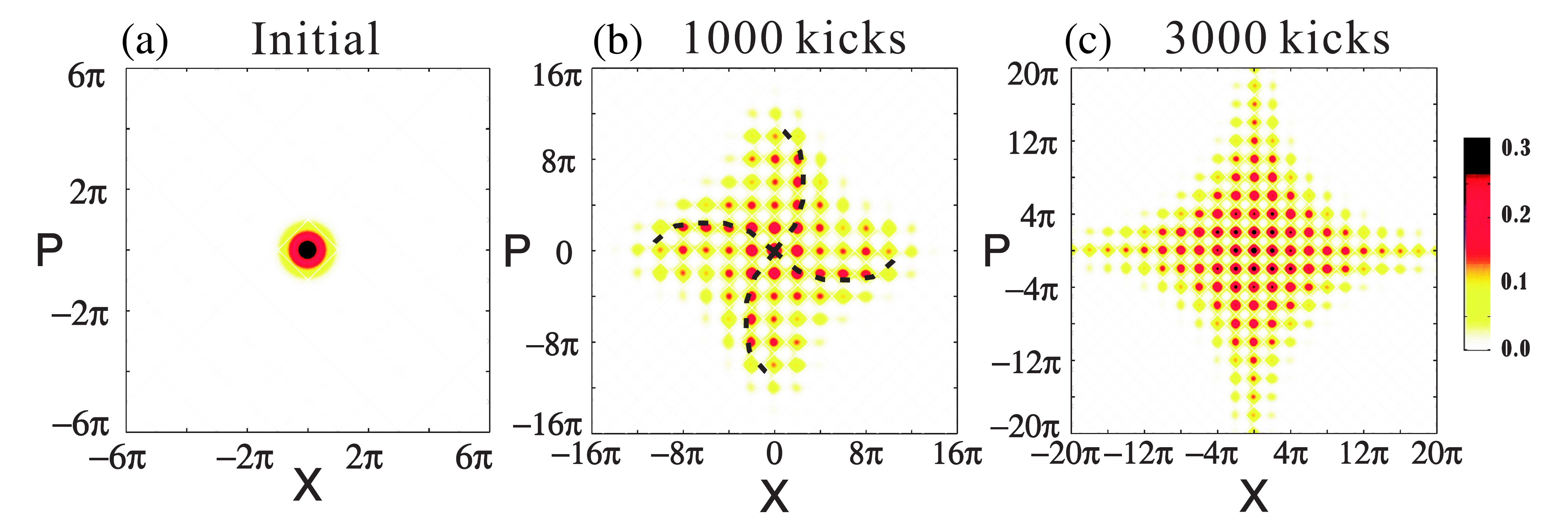}
\caption{\label{fig:disQs}{\bf Dissipative quantum dynamics.} (a) Husimi $Q$-function of the initial state, i.e., the ground state of the harmonic trapping potential. (b) Husimi $Q$-function after $1000$ kicks with the main diffusion path showing a chiral feature (dashed lines). (c) Husimi $Q$-function after $3000$ kicks. Parameters: $K=0.1$ and $\kappa=0.0001$.  The images were taken from Ref.~\cite{liang2018njp}.
}
\end{figure*}

The topology manifests itself also in the dynamics. 
In reality, the driven oscillators are inevitably in contact with the environment resulting in dissipation or decoherence of the quantum system. 
In the Born-Markovian approximation, the dissipative dynamics of the quantum KHO is described by the following master equation \cite{liang2018njp},
\begin{equation}\label{eq:MasterEquation}
\frac{d{\rho}}{dt}=-\frac{i}{\lambda}[H_s(t),\rho]+\kappa(n_0+1)\mathcal{D}[\hat{a}]{\rho}+\kappa n_0\mathcal{D}[\hat{a}^\dagger]{\rho},
\end{equation}
where the Lindblad superoperator is defined by $\mathcal{D}[\hat O]\rho\equiv\hat{O}\rho\hat{O}^\dagger-\frac12(\hat{O}^\dagger\hat{O}\rho+\rho\hat{O}^\dagger\hat{O})
$
and
$\kappa$ characterizes the dissipation rate and $n_0$ is the Bose-Einstein distribution of the thermal bath. It should be emphasized that the above dissipative dynamics of KHO is based on the original full Hamiltonian (\ref{eq:dimensionlessKHO}) without RWA.
To visualize the time evolution of quantum state, we define the Husimi $Q$-function of a given density matrix \cite{knight2004book}
$$
Q(\alpha,\alpha^*)\equiv\frac1\pi\langle\alpha|\rho|\alpha\rangle,
$$
where $|\alpha\rangle$ is the coherent state.  
In Ref.~\cite{liang2018njp}, the time evolution of the Husimi $Q$-function starting from the ground state of the harmonic oscillator was calculated, which is shown in Fig.~\ref{fig:disQs}. 
It is clearly seen that a final steady state with square lattice structure in phase space forms gradually revealing the underlying square structure of Hamiltonian (\ref{eq:Hamiltoniansq}).
Interestingly, the transient state shown in Fig.~\ref{fig:disQs}(b) has no reflection symmetries with respect to $\hat X$ and $\hat P$ albeit the RWA Hamiltonian (\ref{eq:Hamiltoniansq}) has, which results in a chiral feature as marked by the dashed lines along the backbone of the quasiprobability distribution. This chirality is a reflection of the topological property of our system and the noncommutative geometry of the phase space. 
The probability amplitude of a particle appearing at a fixed point (a coherent state in phase space) is the sum of all the possible trajectories. Different from the path integral in 2D real space, each trajectory in phase space associates with a geometric phase due to the noncommutative geometry. The interference of geometric phases breaks the mirror symmetry of phase space. As approaching the stationary state, the chirality disappears in the end since the stationary state should recover the mirror symmetry of the RWA Hamiltonian $(\ref{eq:Hamiltoniansq})$.


In Ref.~\cite{giergiel2019njp}, the Su-Schrieffer-Heeger (SSH) model \cite{su1979prl} was proposed to be realised using bouncing atoms on an oscillating mirror. In the SSH model, the topological phase exhibits zero energy edge states protected by the topology of the bulk. In the same paper, the authors also proposed to simulate the Bose-Hubbard Hamiltonian with repulsive on-site and nearest-neighbor interactions, where the topological Haldane insulator phase \cite{dalla2006prl} may exist. But in order to understand the strongly correlated phenomena, we need to provide a general way to deal with the interaction term in Hamiltonian (\ref{ManybodyH}), which is the centre topic in the next section.

\subsection{Interactions}\label{sec:i}
Until now, we have neglected the interaction terms in the original Hamiltonian (\ref{ManybodyH}). Including the interactions between particles, the total RWA Hamiltonian can be written as
\begin{eqnarray}\label{RWAH}
H_{RWA}^T &=&\sum_{i} H_{RWA}(\hat{X}_i,\hat{P}_i)+ \ \sum_{i<j}U(\hat{X}_i,\hat{P}_i;\hat{X}_j,\hat{P}_j).
\end{eqnarray}
Here, $H_{RWA}(\hat{X}_i,\hat{P}_i)$ is the single-particle RWA Hamiltonian discussed above. The interaction term $U(\hat{X}_i,\hat{P}_i;\hat{X}_j,\hat{P}_j)$ is the RWA part of the transformed real space interaction $V(\hat x_i-\hat x_j)$ in the rotating frame. In general, the interaction term $U(\hat{X}_i,\hat{P}_i;\hat{X}_j,\hat{P}_j)$ is defined in the phase space of the rotating frame and depends on both coordinates and momenta of two particles. Below, we discuss the explicit form of $U(\hat{X}_i,\hat{P}_i;\hat{X}_j,\hat{P}_j)$ both in the classical limit and in quantum mechanics.

\begin{figure}
\centerline{\includegraphics[width=\linewidth]{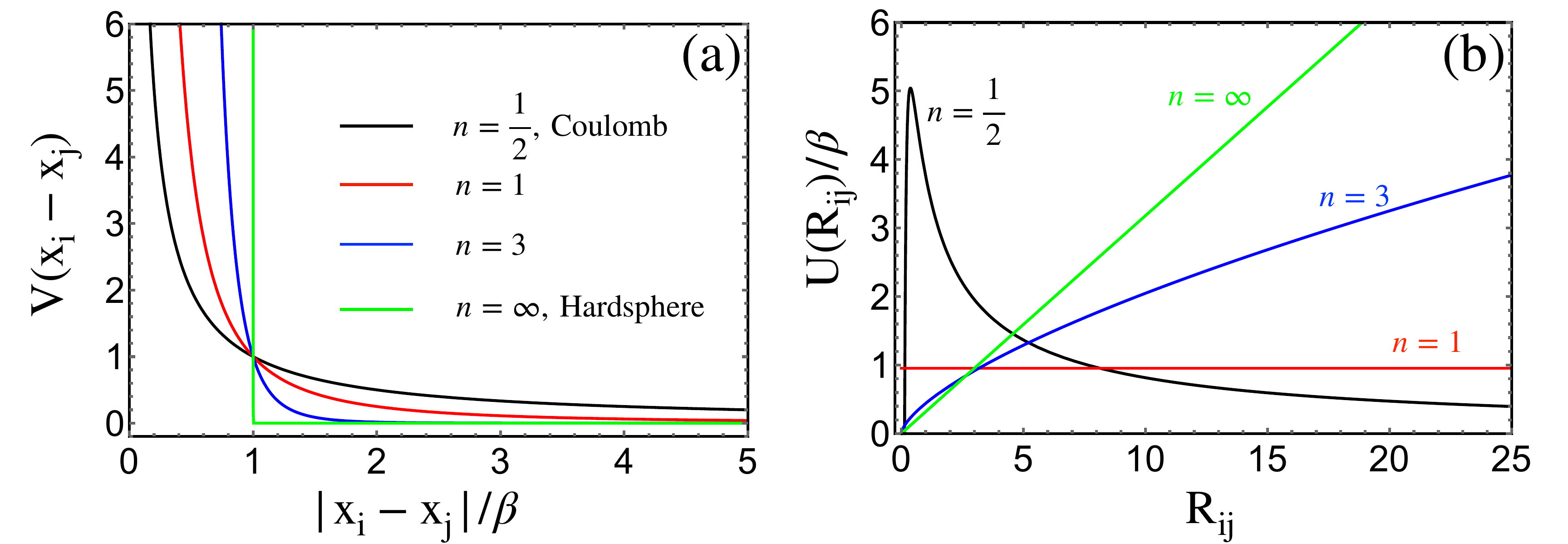}} 
\caption{{\bf Classical phase space interactions.} (a)
Real space power-law interactions. The coloured dashed curves are the example plots for $n=1/2$
(Coulomb interaction), $n=1$, $n=3$ and $n=\infty$ (Hard-sphere
interaction). {(b)} Phase space interactions
corresponding to the power-law potentials shown in figure
(a). Parameter: $\beta=0.01$.
}\label{fig RWAU}
\end{figure}
%

In classical dynamics, we can omit the hats of operators in the Hamiltonian (\ref{RWAH}). For a given real space interaction potential $V(x_i-x_j)$, a general form of $U(X_i,P_i;X_j,P_j)$ has been found in Ref.~\cite{guo2016pra}
\begin{eqnarray}\label{RWA-VR}
U(R_{ij})&=&\int_{-\infty}^{+\infty}dq V_q J_0(qR_{ij})=\frac{1}{2\pi}\int_0^{2\pi}V(R_{ij}\sin\tau)d\tau
\end{eqnarray}
with the defined phase space distance $R_{ij}\equiv\sqrt{(X_i-X_j)^2+(P_i-P_j)^2}$. Here, $V_q$ is the Fourier coefficient $
V_q=(2\pi)^{-1}\int_{-\infty}^{+\infty}dx V(x) \exp(-iqx)$, and
 $J_0(\bullet)$ is the Bessel function of zeroth order. The second equality comes from
the integral representation of the Bessel function, i.e.,
$J_0(x)=(2\pi)^{-1}\int_{-\pi}^{\pi}\exp(-ix\sin\tau)d\tau$, and shows that   $U(R_{ij})$  is in fact the time average of the original
interaction $V(x_i-x_j)$ over the oscillation period. Since $U(R_{ij})$ is defined in the phase space of
the rotating frame and it is only a function of  the phase space
distance $R_{ij}$, we call it the \textit{phase space interaction}.
However, there is a
divergence problem in calculating the phase space interaction $U(R_{ij})$ for, e.g.,
the Coulomb interaction. In Ref.~\cite{guo2016pra}, the origin of divergence was analysed and the correct $U(R_{ij})$  was obtained by introducing a renormalization procedure. For instance, the phase space interaction of the contact interaction in real space $V(x_i-x_j)=\beta\delta(x_i-x_j)$, which is
used to describe the effective interaction between neutral ultra
cold atoms, is given by
$
U(R_{ij})=\pi^{-1}\beta/R_{ij}.
$
A more general interaction potential form is the power-law interaction potential, i.e.,
$
V(x_i-x_j)={\beta^{2n}}/{|x_i-x_j|^{2n}}
$
with integer and half integer $n\geq 1/2$.
If $n=1/2$, the potential is the Coulomb
potential. If $n\rightarrow \infty$, the potential becomes the hard-sphere
potential with a diameter $\beta$. The renormalized phase space interaction for this power-law interaction is
\begin{eqnarray}\label{urij}
{ U}(R_{ij}) = \left\{ \begin{array}{lll} \frac{2\beta}{\pi
R_{ij}}\ln(\beta^{-1}\gamma R^3_{ij}/2),\ \  & \mbox{
         for \
         $n=\frac{1}{2}$}\\
         \frac{2\beta\gamma^{2n-1}4^{\frac{1}{2n}-1}}{\pi(2n-1)}R^{1-\frac{1}{n}}_{ij},\ \  & \mbox{ for \  $ n =1, \frac{3}{2},2,\frac{5}{2},\cdot\cdot\cdot$}\\
         \frac{\beta}{\pi}R_{ij},\ \  & \mbox{
         for \
         $n\rightarrow\infty$}.
        \end{array} \right.
\end{eqnarray}
Here, $\gamma$ is the collision factor given by $\gamma=(4n-1)^{\frac{1}{2n-1}}$.
%
%
The real space power-law interaction is plotted in Fig.~\ref{fig RWAU}(a) and the corresponding phase space interaction is plotted in Fig.~\ref{fig RWAU}(b). Several interesting features of phase space interaction should be pointed out: 
(1) for the Coulomb potential ($n=1/2$), $U(R_{ij})$ still keeps the form of Coulomb's law, up to a logarithmic prefactor (renormalised ``charge" $\beta$);  
(2) for the special case of $n=1$, $U(R_{ij})$ is a constant meaning there is no effective interaction in phase space;
(3) for the special case of $n>1$, ${U}(R_{ij})$ even grows with $R_{ij}$; 
(4) for the hard-sphere interaction ($n\rightarrow\infty$),  $U(R_{ij})$  increases linearly with phase space distance, reminiscent of the confinement interaction between quarks in QCD.
All these predictions have been justified by simulating the many-body dynamics
numerically in Refs.~\cite{guo2016pra,liang2018njp}.


\begin{figure}\label{fig:PSI}
\centering
\includegraphics[width=\linewidth]{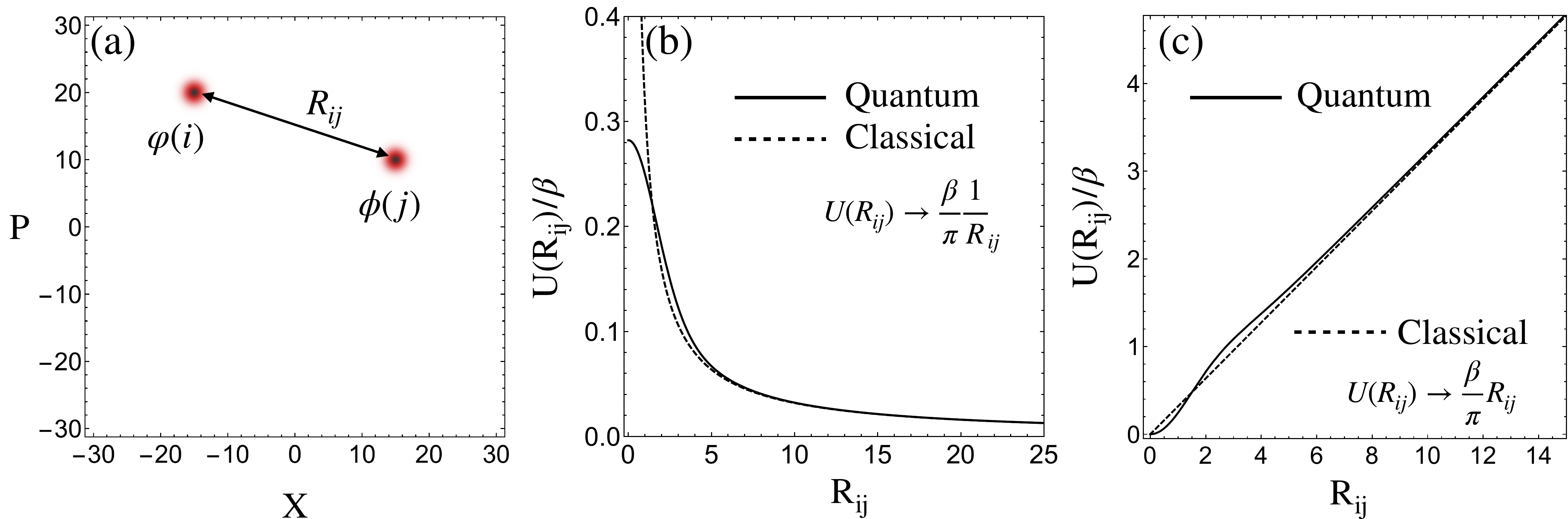}
\caption{\label{fig:PSL-QuantunPSI} {\bf Quantum phase space interactions.} (a) Two coherent states $\varphi(i)$ and $\phi(j)$ in phase space with distance $R_{ij}$ of their centers. (b) Phase space interaction potential $U(R_{ij})$ for the contact interaction. (c) $U(R_{ij})$ for hard-sphere interaction. In both figures (b) and (c), the black solid curve is the quantum interaction ($\lambda=2$) and the blue dashed line is the classical interaction ($\lambda\rightarrow\infty$).
}
\end{figure}

In quantum mechanics, the phase space interaction introduced above for the classical dynamics becomes subtle since the phase space is a noncommutative space. As discussed in Fig.~\ref{fig_QunatumPhase}, the concept of \textit{point} is meaningless in the noncommutative phase space. Instead, it is only allowed to define the coherent state $|\alpha\rangle$ as the \textit{point} in noncommutative geometry. Similarly, the concept of \textit{distance} also needs to be reexamined. It is natural to define the \textit{phase space
distance operator} by
\bea
\hat{R}_{ij}\equiv\sqrt{
(\hat{X}_i-\hat{X}_j)^2+(\hat{P}_i-\hat{P}_j)^2}.
\eea
From the noncommutative relationship $[X_i,P_j]=i\lambda\delta_{ij}$, we have $[\Delta \hat{X}_{ij},\Delta \hat{P}_{ij}]=i(2\lambda)$
where the operators $\Delta \hat{X}_{ij}\equiv\hat{X}_i-\hat{X}_j$, $\Delta
\hat{P}_{ij}\equiv\hat{P}_i-\hat{P}_j$ are the relative displacement of two particles in phase space. Reminiscent of the Hamiltonian operator of a harmonic
oscillator, we see that  the phase space distance is a quantized variable with a lower limit $\sqrt{2\lambda}$. In Ref.~\cite{liang2018njp}, a compact
expression for the phase space interaction in the operator form was given by
\begin{eqnarray}\label{RWA-VLa}
U(\hat{R}_{ij})=\int_{-\infty}^{+\infty}dq V_q e^{-\frac{\lambda
q^2}{2}} L_{\frac{1}{4\lambda}\hat{R}^2_{ij}-\frac{1}{2}}(\lambda q^2).
\end{eqnarray}
Here, $V_q=(2\pi)^{-1}\int_{-\infty}^{+\infty}dxV(x)\exp(-iqx)$ is the Fourier coefficient of the real space interaction $V(x_i-x_j)$ and $L_{n}(\bullet)$ is the Laguerre polynomials. 

If the quantum states of two particles are two wave packet $\phi(i)$ and $\varphi(j)$ as shown in Fig.~\ref{fig:PSL-QuantunPSI}(a), their averaged phase space interaction is given by
\bea\label{Ud}
U_{ij}&=&\langle\varphi(i)\phi(j)|U(\hat{R}_{ij})|\varphi(i)\phi(j)\rangle.
\eea
The explicit expressions of $U_{ij}$ for the two displaced coherent states were calculated in Ref.~\cite{liang2018njp}. For the contact
interaction $V(x_1-x_2)=\beta\delta(x_1-x_2)$, the resultant $U_{ij}$ is a function of the phase space distance between the centers of two coherent states $R_{ij}$, i.e., $U_{ij}=U(R_{ij})$ with
\begin{eqnarray}\label{UcRcontact}
U(R_{ij})&=&\frac{\beta}{\sqrt{2\pi\lambda}}\exp\Big(-\frac{R^2_{ij}}{4\lambda}\Big)\
I_0\Big(\frac{R^2_{ij}}{4\lambda}\Big)
\end{eqnarray}
Here,  $I_0(\bullet)$ is the zeroth order modified Bessel function
of the first kind. For the hard-sphere interaction, i.e., $V(x_i-x_j)={\beta^{2n}}/{|x_i-x_j|^{2n}}$ for
$n\rightarrow \infty$, the result is 
\begin{eqnarray}\label{UceHardcore}
U(R_{ij})&\approx&\frac{4\beta\sqrt{\lambda}}{\pi}\sum_{m=0}^\infty\,\frac{\sqrt{2m+3/2}}{(2m+1)!}\,\exp\Big(-\frac{R^2_{ij}}{4\lambda}\Big)
\Big(\frac{R^2_{ij}}{4\lambda}\Big)^{2m+1}.
\end{eqnarray}
In Fig.~\ref{fig:PSL-QuantunPSI}, the $U(R_{ij})$ as {\color{blue}a} function of $R_{ij}$ and its
long-range asymptotic behaviour are plotted. We see that a point-like contact
interaction indeed becomes a Coulomb-like interaction $U(R_{ij}) \sim{\pi^{-1} }\beta/R_{ij}$ in the long-distance limit $R_{ij}\gg 2\sqrt{\lambda}$, and the phase space interaction of the hard-sphere potential becomes linear in the long-distance limit $U(R_{ij})\rightarrow {\beta}{\pi^{-1}}R_{ij}$, which are consistent with the pure classical analysis. 


\subsection{Mott insulator}\label{sec:mi}

With the theory of phase space interaction, we can extend the tight-binding model (\ref{TBM}) for free particles to the many-body Hamiltonian with interactions.
Assuming the particles are spinless Bosons, we can write the Floquet many-body Hamiltonian for the $l$-th band directly
\bea\label{TBMU}
H_{TB}&\approx& E_l\sum_{i}\hat n_{i,l}+J_l\sum_{\langle i,j\rangle}\hat a^\dagger_{i,l}\hat a_{j,l}\nonumber\\
&&+\sum_i\frac{1}{2}U_{ii,l}\hat n_{i,l}(\hat n_{i,l}-1)+\frac{1}{2}\sum_{i\neq j}U_{ij,l}\hat n_{i,l}\hat n_{j,l}.
\eea 
Here, $J_l$ is the tunnelling rate between nearest lattice sites denoted by $\langle i,j\rangle$, and $\hat n_{i,l}=\hat a^\dagger_{i,l}\hat a_{i,l}$ is the number operator occupying the Fock space $|\cdots,n_{i,l},\cdots\rangle$ in the Wannier representation. Model Hamiltonian (\ref{TBMU}) is valid if the interaction $U_{ij,l}$ is weak compared to the quasienergy gaps. If the phase space interaction $U_{ij,l}$ does not have compact form as given by Eq.~(\ref{UcRcontact}) or it is difficult to obtain an analytical form, 
we can calculate it numerically from Eq.~(\ref{Ud}) or refer to the original expression in the laboratory frame
\bea\label{Uijt}
U_{ij,l}=\int^T_0\mathrm{d} t \langle \phi_{i,l}(x_m,t)\phi_{j,l}(x_n,t)|V(x_m-x_n)|\phi_{i,l}(x_m,t)\phi_{j,l}(x_n,t) \rangle,\nonumber
\eea
where $ \phi_{i,l}(x_m,t)$ represents the $l$-th Wannier wave function on the phase space lattice site $i$ for the $m$-th particle in the laboratory frame. For the contact interaction $V(x_m-x_n)=g_0\delta(x_m-x_n)$, we have $U_{ij,l}=g_0\int^T_0\mathrm{d} t \int \mathrm{d} x |\phi_{i,l}(x,t)|^2|\phi_{j,l}(x,t)|^2$. This is the method adopted by Sacha {\it et al}  in their study \cite{sacha2015scirep,sacha2015pra,giergiel2019njp,kosior2018pra,giergiel2018pra,giergiel2019prb,sacha2017prb,sacha2018prl}.

In Ref.~\cite{sacha2015scirep}, Sacha {\it et al} discussed the Mott transition using the model of bouncing atoms on an oscillating mirror. In the limit of free interaction, the ground state of many-body Hamiltonian (\ref{TBMU}) is a superfluid state with long-time phase coherence. However, for sufficiently strong repulsive contact interaction ($g_0>0$) such that $U_{ii,l}\ll N|J_l|/n$ (neglecting off-site interaction terms), the ground state of (\ref{TBMU}) is a single Fock state with well defined numbers
of atoms occupying each Wannier state, i.e., $|N/n,N/n,\cdots,N/n\rangle$ with $N$ denoting the total number of particles. As a result, a gap opens between the ground state level and excited levels, thus consequently a Mott insulator phase is achieved. Sacha {\it et al} also investigated the model (\ref{TBMU}) in two dimensional space by extending atoms bouncing between two orthogonal harmonically oscillating mirrors \cite{giergiel2019prb,giergiel2018prl}. By including off-site interactions in (\ref{TBMU}), more intriguing condensed matter phenomena, e.g., the topological Haldane phase mentioned in the end of Sec.~\ref{sec:top} ,  are possible to be explored \cite{giergiel2019njp}.

\subsection{Heisenberg model}\label{sec:hm}

If the two particles are indistinguishable and have spins (i.e., Fermions with half spin $\hat s_i$), their spatial state is either antisymmetric or symmetric depending on the symmetry of total spin state, i.e., 
\bea
\psi_{\pm}(i,j)=\frac{1}{\sqrt{2}}\Big[\varphi(i)\phi(j)\pm\phi(i)\varphi(j)\Big].\nonumber
\eea
For singlet (triplet) spin state the wave function is symmetric (antisymmetric). Therefore,
the average phase space interaction potential of the (anti)symmetric state is 
\bea
\langle
\psi_{\pm}(i,j)|U(\hat{R})|\psi_{\pm}(i,j)\rangle
\equiv U_{ij}\pm J_{ij}.\nonumber
\eea
Here, the direct interaction part $U_{ij}$ has been given by Eq.~(\ref{Ud}) while the the exchange interaction part is
\begin{eqnarray}\label{Ue}
J_{ij}&=&\langle\varphi(i)\phi(j)|U(\hat{R})|\phi(i)\varphi(j)\rangle.
\end{eqnarray}
The direct interaction part $U_{ij}$ corresponds to the classical interaction while the exchange interaction part $J_{ij}$ is a
pure quantum effect without classical counterpart, which we call the {\it Floquet exchange interaction energy}.

The Floquet exchange interaction energy of two coherent states was calculated from Eq.~(\ref{Ue}) in Ref.~\cite{liang2018njp}. For the contact interaction, it is found that the exchange interaction is equal to the direct interaction, i.e., 
$
J(R_{ij})=U(R_{ij}).
$
The equality comes from the $\delta$ function modelling the contact interaction and the fact that the spatial antisymmetric state of two particles has zero probability to touch each other. If the interaction potential differs from the $\delta$-function model, the phase space interaction $U_{ij}$ and the Floquet exchange interaction $J_{ij}$ can behave independently. For example, if the two particles are charged and confined in quasi-1D ion trap, the direct interaction and the Floquet exchange interaction were calculated in Ref.~\cite{liang2018arxiv}, i.e., 
$U(R_{ij})\propto 1/R_{ij}$ and $J(R_{ij})\propto \hbar/R^3_{ij}$. 
In this case, the exchange interaction decays in the form of inverse-cube law and is a quantum effect due to the prefactor $\hbar$. One should always remember that we are studying the stroboscopic dynamics of a periodically driven system. The two particles collide with each other during every stroboscopic time step, and the exchange interaction comes from the overlap of their wave functions during the collision process. For the hard-sphere interaction, the particles cannot cross each other \cite{liang2018njp}, and thus the Floquet exchange interaction is zero $J_{ij}=0$.

\begin{figure}
\centerline{\includegraphics[width=\linewidth]{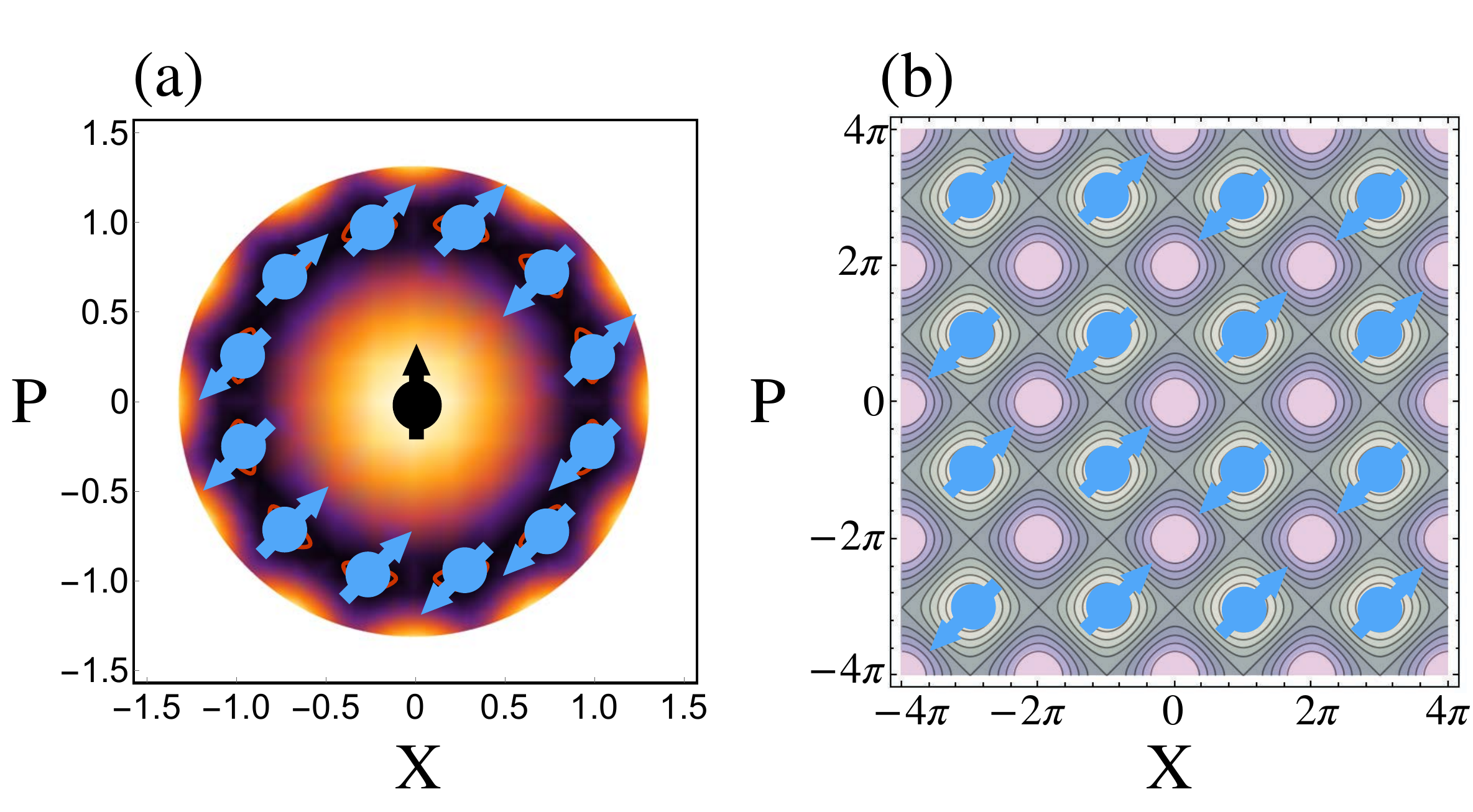}}
\caption{\label{Fig:Heisenberg}{\bf Heisenberg models.} (a) 1D Heisenberg model of periodically arranged spins (blue) on the $\mathbb{Z}_n$ phase space lattice sites interacting with a central spin (black) in the origin. (b) 2D Heisenberg model of spins occupying one sublattice of the square lattice in phase space.}
\end{figure}

If the particles are tightly bound at their equilibrium points by the phase space lattice, the direct phase space interaction $U(R_{ij})$ does not play a role in the dynamics. The dynamics of the spins is determined by the Floquet exchange interaction and is described by the Heisenberg model  with isotropic spin-spin interaction, i.e.,
 \bea
H_{ex}=-\frac 1 4 \sum_{i\neq j} J(R_{ij}) \hat{S}_i \cdot \hat{S}_j+\mathrm{constant}
\eea
In Fig.~\ref{Fig:Heisenberg}, we show the Heisenberg models in the $\mathbb{Z}_n$ circular lattice and square lattice in phase space. The operator $\hat S_i\equiv \sum_m \hat s_{im}$ is the total spin operator on the $i$-th site of phase space lattice since more than one spins can be confined in one lattice site. For the $\mathbb{Z}_n$ circular lattice, the origin point $(0,0)$ in phase space is also a stable point, which can hold a central spin interacting with the spins on the $\mathbb{Z}_n$ lattice. Therefore, the corresponding Heisenberg model up to a constant is
 \bea
H_{ex}=-\frac 1 4 \sum^n_{i\neq j} J(R_{ij}) \hat{S}_i \cdot \hat{S}_j-\frac 1 4 \sum^n_{i=1} J(R_{i0}) \hat{S}_0 \cdot \hat{S}_i,
\eea
where $\hat S_{i(j)}$ with $i(j)=1,2,\cdots,n$ represents the spins on the $\mathbb{Z}_n$ lattice sites and $\hat S_0$ is the central spin. This model is closely related to the central spin models \cite{claeys2018prl,ashida2019prl}.
The famous Mermin-Wagner theorem claims that a 1D or 2D isotropic Heisenberg model with finite-range exchange interaction can be neither ferromagnetic nor antiferromagnetic \cite{mermin1966prl} at any nonzero temperature, which clearly excludes a variety of types of long-range ordering in low dimensions.  Here in our model, the Flquet exchange interaction from point-like contact interaction has a Coulomb-like long-range behaviour. Therefore, the phase space lattices provide a possible platform to test the Mermin-Wagner theorem and search for other new phenomena with long-range interactions such as causality and quantum criticality \cite{hauke2013prl,richerme2014nature,gong2014prl,metivier2014prl,foss2015prl,maghrebi2016prl,buyskikh2016pra}, nonlocal order \cite{dalla2006prl,berg2008prb,endres2011science}, etc.

\subsection{Disorder $\&$ Localisation}\label{sec:dl}


By introducing a random temporal perturbation satisfying $H'(t+nT_d)=H'(t)$ with $T_d=2\pi/\omega_d$ to match the $n:1$ nonlinear resonance \cite{sacha2015scirep,sacha2017rpp}, a random on-site potential in the tight-binding model can be produced,
\be\label{eq:altb}
H_{\mathrm{eff}}\approx -\frac12\sum_{j=1}^n(J_j\hat a^\dagger_{j+1}\hat a_j+\mathrm{c.c.})+\sum_{j=1}^n\epsilon_j\hat a_j^\dagger \hat a_j,
\ee
where
\be\label{onsite}
\epsilon_j=\int_0^{n2\pi/\omega_d}dt\int_0^\infty dz H'(t)|\phi_j|^2
\ee
can be modulated by properly choosing the fluctuating function in $H'(t)$. For the $\mathbb{Z}_n$ lattice, Eq.~(\ref{eq:altb}) is the 1D Anderson model on a ring. In the 1D Anderson model, the localization always happens no matter how weak the disorder is. In the phase space picture, the localised state is the superposition of the wave-packets around the  $\mathbb{Z}_n$ lattice site and with probability exponentially suppressed as the phase space distance increases. In the laboratory frame, the interpretation is that the probability density at a fixed position in the configuration space is localised exponentially around a certain moment of time \cite{sacha2015scirep}, which is coined as the {\it Anderson localisation in time domain}.

A different but related scheme without utilizing classical nonlinear resonances is presented in \cite{sacha2016pra} by Sacha and Delande. In this case the authors considered the problem of a single particle moving on a ring geometry under the action of a periodic force which however fluctuates randomly in time. The model is described by the following classical Hamiltonian,
\bea
H={(p-\alpha)^2}/{2}+V_0g(\theta)f(t),\nonumber
\eea
where the position of the particle on the ring is denoted by the angle variable $\theta$ and its canonical momentum by $p$. The Anderson problem is considered by introducing random disorder in space whereas here it is in time. Specifically, the time-dependent force consists of independent random Fourier components, e. g., 
$
f(t)=\sum\limits_{k=-\infty}^\infty f_ke^{ik\omega_d}
$
with $f_k=f_{-k}^*$ independent random variables and $\omega_d$ the driving frequency. Note that the spatial potential $g(\theta)$ is not necessarily disordered but any regular function. 
Inside the classical invariant torus one can employ the secular approximation and eliminate the fast oscillating part of the motion and arrive at an effective Hamiltonian describing the slow motion,
\bea\label{Hdisorder}
H\approx\frac{P^2}{2}+V_0\sum\limits_{k\ne 0}g_k f_{-k} e^{ik\Theta}+{\rm constant}.
\eea
Here $P$ and $\Theta$ are canonical variables of the slow motion and they are related to the original ones by the transformation $\Theta=\theta-\omega_d t$ and $ P=p-\alpha-\omega_d$. Assume the disordered potential satisfies the normal distribution $|g_kf_{-k}|\propto e^{-k^2/(2k_0^2)}$ with correlation length $\sqrt{2}/k_0$ and $\mathrm{Arg}(f_{-k})$ are uniform random numbers in the interval $[0,2\pi)$. With a suitable choice of $V_0$ and $k_0$, the eigenstates of 
(\ref{Hdisorder}) are Anderson localized even for energies higher than the standard deviation of the disordered
potential. 
The Anderson localization in time crystals is revealed by measuring the probability that the particle passes a fixed position, which is localized exponentially around a certain moment of time. A possible experimental scheme for realising such Anderson localization is an electron in a Rydberg atom perturbed by a fluctuating microwave field \cite{giergiel2017pra}. The Anderson model was later extended to higher dimensions by considering particles moving in a three dimensional torus \cite{sacha2017prl230404}.

It is quite natural to combine interactions between particles
with at the same time added temporal disorder, which
extends the analysis of the Anderson localisation to interacting particle
systems, i.e., the many-body localisation (MBL). In Ref.~\cite{sacha2017prb}, Sacha {\it et al} proposed the following Bose-Hubbard Hamiltonian with on-site disorder
\bea\label{MBHdis}
H_{\mathrm{eff}}\approx -\frac J 2\sum_{j=1}^n(\hat a^\dagger_{j+1}\hat a_j+\mathrm{c.c.})+\frac 1 2 \sum^n_{i,j}U_{ij}\hat a_i^\dagger \hat a_i\hat a_j^\dagger \hat a_j+\sum_{j=1}^n\epsilon_j\hat a_j^\dagger \hat a_j,
\eea
where the interaction $U_{ij}$ and on-site disorder $\epsilon_i$ are given by Eq.~(\ref{onsite}) and Eq.~(\ref{Uijt}) respectively. The Hamiltonian (\ref{MBHdis})  is valid provided the interaction energy
$NU_{ij}$ ($N$ is a total number of bosons) and the disorder $\epsilon_j$
are much smaller than the energy gap. Although the disorder breaks the translational invariance of the phase space lattice, the system still possesses
Floquet eigenstates with longer period $2\pi n/\omega_d$ equal to the period of disorder, i.e., $H'(t+nT_d)=H'(t)$. The MBL in disordered Hubbard model has been observed in the experiment \cite{schreiber2015science}. It is believed that MBL is constrained to systems with
short-range interactions \cite{yao2014prl} but the long-range interaction alone
does not exclude MBL \cite{nandkishore2017prx}. The existence of MBL with long-range interactions is an important but unexplored problem. In the experiments with cold atoms, the interaction is short range since the atoms are neutral in general. The many-body Hamiltonian (\ref{MBHdis}) from subresonantly driven protocol may present a new approach to solve this problem, as the long-range interaction $U_{ij}$ can be generated from contact interaction directly.

By modulating the strength of contact interaction with a proper temporal disorder, the particles can have an effective disordered potential in phase space. In such way, a two-particle molecule bound by Anderson localisation is created from the periodic driving \cite{sacha2018prl}.  



\begin{figure}\label{fig:PSI}
\centering
\includegraphics[width=0.7\linewidth]{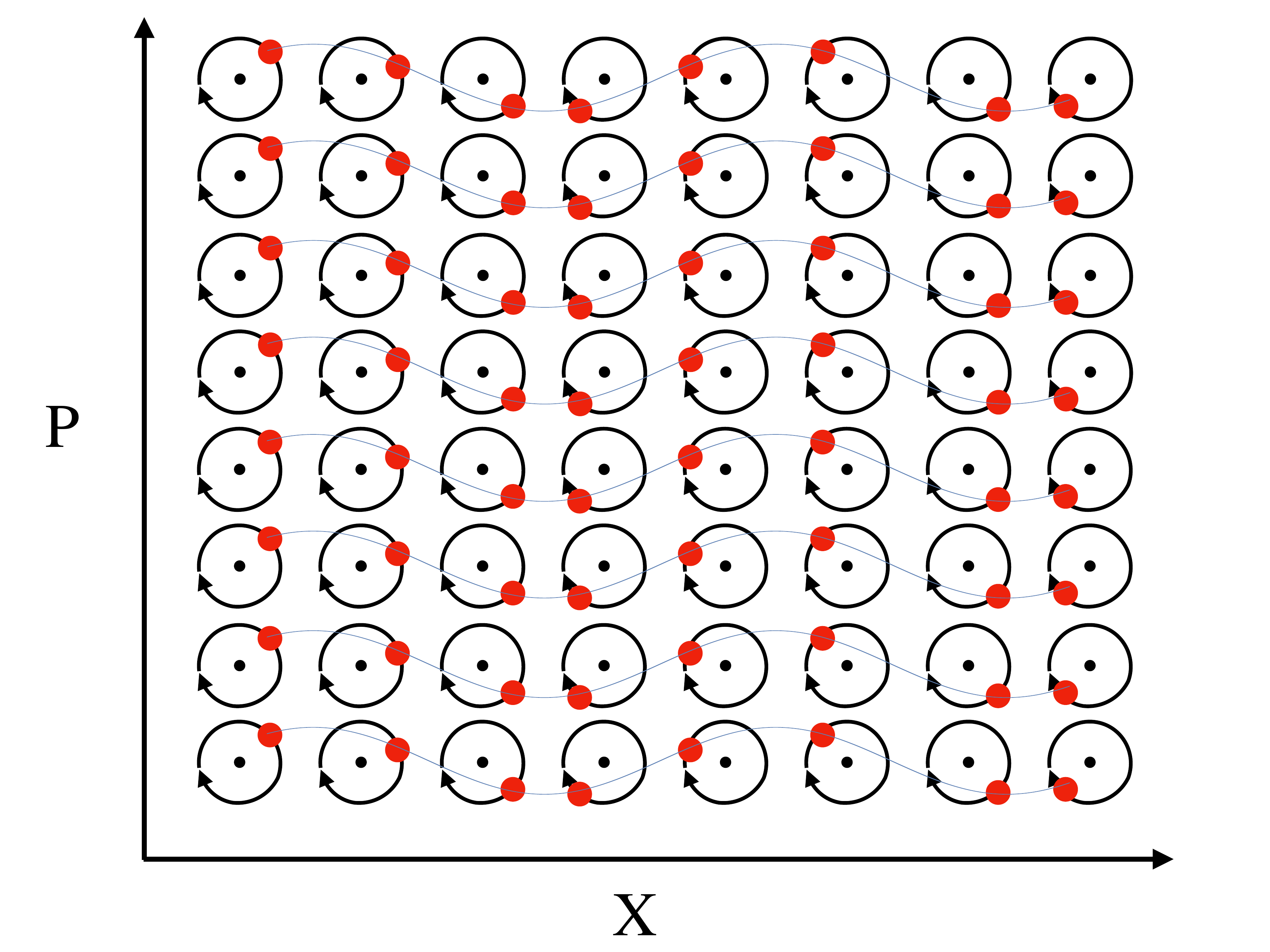}
\caption{\label{Fig-Future-waves} {\bf Phase space lattice waves.} The atoms (red dots) rotate around their equilibrium points unidirectionally, i.e., clockwise or anticlockwise, depending on the extremal properties  (maximum or minimum) of the equilibrium points. The collective motions may arise from the phase space interaction between particles, which are the lattice waves in phase space.
}
\end{figure}

\section{Summary and Outlook}\label{Sec:SummaryandOutlook}

We started from Wilczek's time crystals breaking the CTTS of a static Hamiltonian \cite{wilczeck2012prl-1,wilczeck2012prl-2}, which has been disproved to exist  in equilibrium systems \cite{bruno2013prlcomment,Nozieres2013epl,bruno2013prl,WO2015prl,khemani2017prb}. However, in periodically driven (Floquet) systems, the time crystal behaviour does exist by breaking the DTTS \cite{sacha2015pra,khemani2016prl,else2016prl,von2016prb,yao2017prl}, namely, the system responds at a fraction $\omega_d/n$ of the original driving frequency. Such Floquet time crystal is a robust subharmonic mode oscillating in the laboratory frame with time period $nT_d$, which is $n$ times the driving period $T_d=2\pi/\omega_d$. Most works so far on time crystals are interested in stabilising the subharmonic mode via various  mechanisms like MBL in spin chain models. This review paper, however, focuses on the many-body dynamics of subharmonic modes and the analogous condensed matter phenomena in Floquet time crystals \cite{guo2013prl,sacha2015scirep,guo2016njp,sacha2017prl230404,sacha2017prb,liang2018njp,sacha2018prl}. 
Throughout this review, we choose the phase space lattice picture to view the Flqouet time crystals or the subharmonic modes. Namely, we go to the rotating frame at the frequency of subharmonic modes, and place all the subharmonic modes in the phase space of the rotating frame according to their oscillating amplitudes and phases (or the action and angle variables). The resulting arrangement shows a periodic crystal structure in phase space, e.g., a  circular $\mathbb{Z}_n$ lattice and a 2D square lattice.  The phase space lattice picture in the rotating frame \cite{guo2013prl,guo2016njp,liang2018njp}, which is equivalent to the time crystal picture in the laboratory frame \cite{sacha2015scirep,sacha2018prl}, is convenient for developing an analogous condensed matter theory as in solid state physics. We reviewed the single-particle band theory for phase space lattice \cite{guo2013prl,guo2016njp} and the topological phenomena due to the noncommutative geometry of phase space \cite{liang2018njp}. We also discussed the effective interaction between two particles in phase space \cite{guo2016pra,liang2018njp} beyond the single-particle picture. At this setting, the many-body Hamiltonian like  the Hubbard model \cite{sacha2015scirep,sacha2018prl} and  the Heisenberg model \cite{liang2018njp} can be realised in phase space lattice. By adding temporal disorder to the system, the Anderson model \cite{sacha2015scirep,sacha2016pra,giergiel2017pra,sacha2017prl230404} and the disordered Bose-Hubbard model \cite{sacha2017prb,sacha2018prl} are also possible to be created in time crystals.

It is difficult to predict the research directions in the future although there are many unexplored topics in this new research field. Except mimicking the existing models in condensed matter physics, it would be more interesting to find something new in phase space lattice. 
For example, a {\it phase space crystal} should exist if the phase space lattice is strong enough to confine the atoms near its stable points. In this case, it is interesting to study the collective motion of atoms near their equilibrium points, which may form {\it phase space lattice waves} as indicated in Fig.~\ref{Fig-Future-waves}. The lattice waves in phase space differ from the lattice waves in real space at a fundamental level that they are waves in noncommutative space. Each atom actually only rotates around its equilibrium position unidirectionally, i.e., either clockwise or anticlockwise depending on the extremal properties  (minimum or maximum) of equilibrium points. Therefore, we conjecture the phase space lattice waves could have some topological properties, i.e., the nontrivial edge states \cite{guo2020}.

\begin{figure}\label{fig:PSI}
\centering
\includegraphics[width=0.8\linewidth]{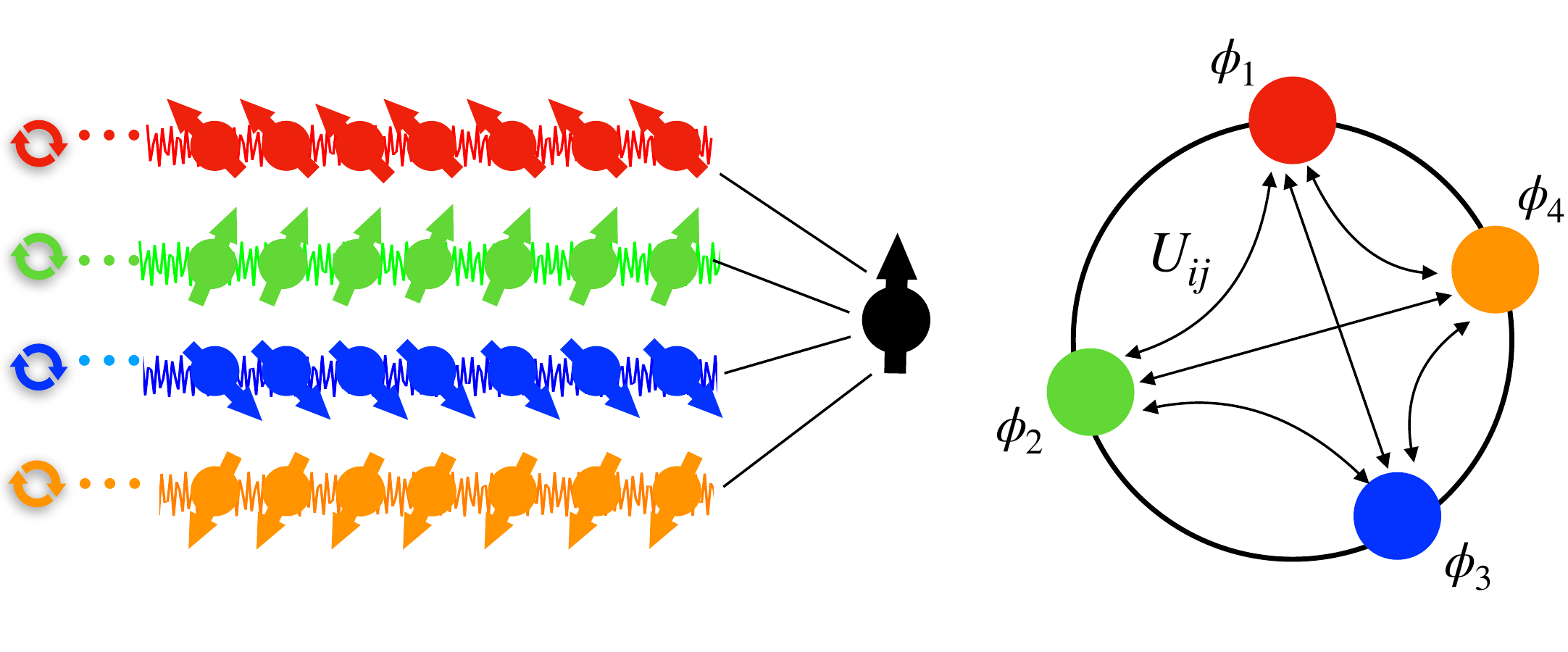}
\caption{ \label{Fig-Future-spins}{\bf A gas of Floquet time crystals with spin chains.} (Left) Four periodically driven spin chains (arrows with different colours) couple to an auxiliary spin (black arrow) via their rightmost spins. The circular arrows on the leftmost sides represent the periodic driving field and the noisy curves threading the spin chains represent the disorder to stabilise the subharmonic modes. (Right)  Any two subharmonic modes of spin chains (coloured rounds) have the same effective interaction form $U_{ij}$, which only depends on their oscillating phase $\phi_i$ and $\phi_j$. The ensemble of subharmonic modes form a gas of Floquet time crystals.
}
\end{figure}

Until now, the condensed matter phenomena in time crystals are only studied with the models of periodically driven quantum oscillators (i.e., driven quantum gas).  One may ask if it is possible to develop a similar theory with periodically driven spin chain models. However, one should notice there is an important difference between these two models, i.e, each subharmonic mode of the driven oscillator can be a single-particle state while the subharmonic mode in the driven spin chain model is already a many-body state of spins. Nevertheless, we can investigate the many-body dynamics of subharmonic modes with spin models. As illustrated in Fig.~\ref{Fig-Future-spins}, imagine there are several periodically driven spin chains interacting with an auxiliary centre spin on an equal footing. The subharmonic modes oscillate with frequency in the form of $\cos(\omega_dt/n+\phi_i)$, where $\phi_i$ is the oscillating phase of each subharmonic mode. The effective interaction between any two subharmonic modes should have the same form depending only on their oscillating phases. Then, the many-body dynamics of subharmonic modes can be described by the general Hamiltonian (\ref{RWAH}). In this way, the ensemble of subharmonic modes can be cast as a {\it gas of Floquet time crystals}. Some novel phenomena may arise in this model. For example, if the effective interaction is negligible, the time symmetry breaking processes of different spin chains are independent. However, if the interaction with centre spin is strong enough, the time symmetry breaking processes of different spin chains may affect each other and form a crystal in the oscillating phase $\phi_i$ of time crystals, namely, from a gas-like state to a solid-like state.

The effects of the non-RWA Hamiltonian are also not fully understood yet.
The phase space lattice is an emergent property of the RWA Hamiltonian, which is the leading Floquet-Magnus term in the rotating frame ${H}_{RWA}={H}^{(0)}_F$.  The higher-order Floquet-Magnus terms  from the non-RWA Hamiltonian [e.g., Eq.~(\ref{hnonrwa})] deteriorates the discrete lattice symmetry of the RWA Hamiltonian [e.g., Eq.~(\ref{H})]. One can ask how the quasienergy band structure changes in the presence of non-RWA terms. In Ref.~\cite{guo2013prl}, it has been found that the band structure of the $\mathbb{Z}_n$ lattice (\ref{HZn}) is very robust for a large region beyond the RWA regime, but the tunnelling rate $J_l$ shows some sharp
peaks at some detuning parameter, which has been identified as the resonant transitions induced by the fast non-RWA oscillating terms disregarded 
in the RWA \cite{peano2012prl}. In general, the current approach only applies for the stroboscopic dynamics of Floquet systems. The study on the non-RWA Hamiltonian describing the micromotions of system \cite{bukov2015aip} is an important research direction in the future. Another interesting and fundamental research direction is the classification of possible space-time groups for time crystals, which is analogous to the classification of space groups for static crystals. In Ref.~\cite{xu2018prl}, a complete classification of the 13 space-time groups in one-plus-one dimension (1+1D) is performed, and $275$ spacetime groups are classified in 2+1D. Such study serves as the starting point for future research on the topological properties of time crystals.

\section*{Acknowledgements}
The authors thank Prof. Florian Marquardt for useful discussions; and Prof. Krzysztof Sacha for helpful comments.

\appendix


\providecommand{\newblock}{}

\end{document}